\documentclass[twocolumn,prd,showpacs,amssymb,amsmath,eqsecnum]{revtex4}
\usepackage{bm}
\usepackage{graphicx}
\usepackage{accents}

\newcommand{\etal}{\textit{et al.}}

\newcommand{\beq}{\begin{equation}}
\newcommand{\eeq}{\end{equation}}
\newcommand{\ein}{{\epsilon_{\rm in}}}
\newcommand{\eout}{{\epsilon_{\rm out}}}
\newcommand{\bein}{{\bm{\epsilon}_{\rm in}}}
\newcommand{\beout}{{\bm{\epsilon}_{\rm out}}}

\newcommand{\beoutsup}[1]{{\bm{\epsilon}_{\rm out}^{#1}}}
\newcommand{\einsup}[1]{{{\epsilon}_{\rm in}^{#1}}}
\newcommand{\eoutsup}[1]{{{\epsilon}_{\rm out}^{#1}}}
\newcommand{\rhat}{\hat{\bf r}}
\renewcommand{\Re}{\mathop{\rm Re}}
\renewcommand{\Im}{\mathop{\rm Im}}
\newcommand{\kappasub}[1]{\ensuremath{{\kappa_{_{#1}}}}}
\newcommand{\kappasubsq}[1]{\ensuremath{{\kappa^2_{_{#1}}}}}
\newcommand{\tr}{\mathop{\rm Tr}}

\newcommand{\csq}{\overline{c^2}}
\newcommand{\ssq}{\overline{s^2}}
\newcommand{\cc}{\overline{c}}

\newcommand{\Vo}{\accentset{\circ}{V}}

\begin{document}
\title{Systematic Errors in Cosmic Microwave Background Interferometry}
\author{Emory F. Bunn}
\email{ebunn@richmond.edu}
\affiliation{Physics Department, University of Richmond, Richmond, VA 23173}
\date{\today}

\begin{abstract}
Cosmic microwave background (CMB) polarization observations will require
superb control of systematic errors in order to achieve their full
scientific potential, particularly in the case of attempts to detect
the $B$ modes that may provide a window on inflation.  Interferometry
may be a promising way to achieve these goals.  This paper presents
a formalism
for characterizing the effects of a variety of systematic errors
on interferometric CMB polarization observations, with particular emphasis
on estimates of the $B$-mode power spectrum.  The
most severe errors are those that couple the temperature anisotropy
signal to polarization; such errors include cross-talk within
detectors, misalignment of polarizers, and cross-polarization.
In a $B$ mode experiment, the next most serious category
of errors are those that mix $E$ and $B$ modes, such as gain fluctuations,
pointing errors, and beam shape errors.
The paper also indicates which sources of error may cause circular polarization
(e.g., from foregrounds) to contaminate the cosmologically
interesting linear polarization channels, and conversely whether
monitoring of the circular polarization channels may yield useful
information about the errors themselves.
For all the sources of error considered, estimates of the level of control
that will be required for both $E$ and $B$ mode experiments are provided.
Both experiments that interfere linear polarizations and those that interfere
circular polarizations are considered.  The fact that circular
experiments simultaneously measure both linear polarization Stokes
parameters in each baseline mitigates some sources of error.
\end{abstract}

\maketitle

\section{Introduction}

Cosmic microwave background (CMB) polarimetry is one of the most
exciting frontiers in cosmology.  
CMB polarization has already been detected 
\cite{kovac,wmappol,readhead,leitch,barkats,wmap3pol}, 
and we may expect future instruments to characterize
the polarization signal in much greater detail (e.g., \cite{korotkov}).
In the near future, CMB polarization
data are expected to refine estimates of cosmological parameters
\cite{kinney},
probe the ionization history of the Universe \cite{zal97}
and the details of recombination \cite{peebles},
and measure gravitational lensing due to large-scale structure
\cite{zalsellens}.
Most exciting of all, polarization maps
may provide a direct probe of an inflationary epoch in the extremely
early Universe by detecting the signature of primordial
gravitational radiation \cite{zalsel,selzal,kkslett,kks}.

\begin{figure*}
\includegraphics[width=3in]{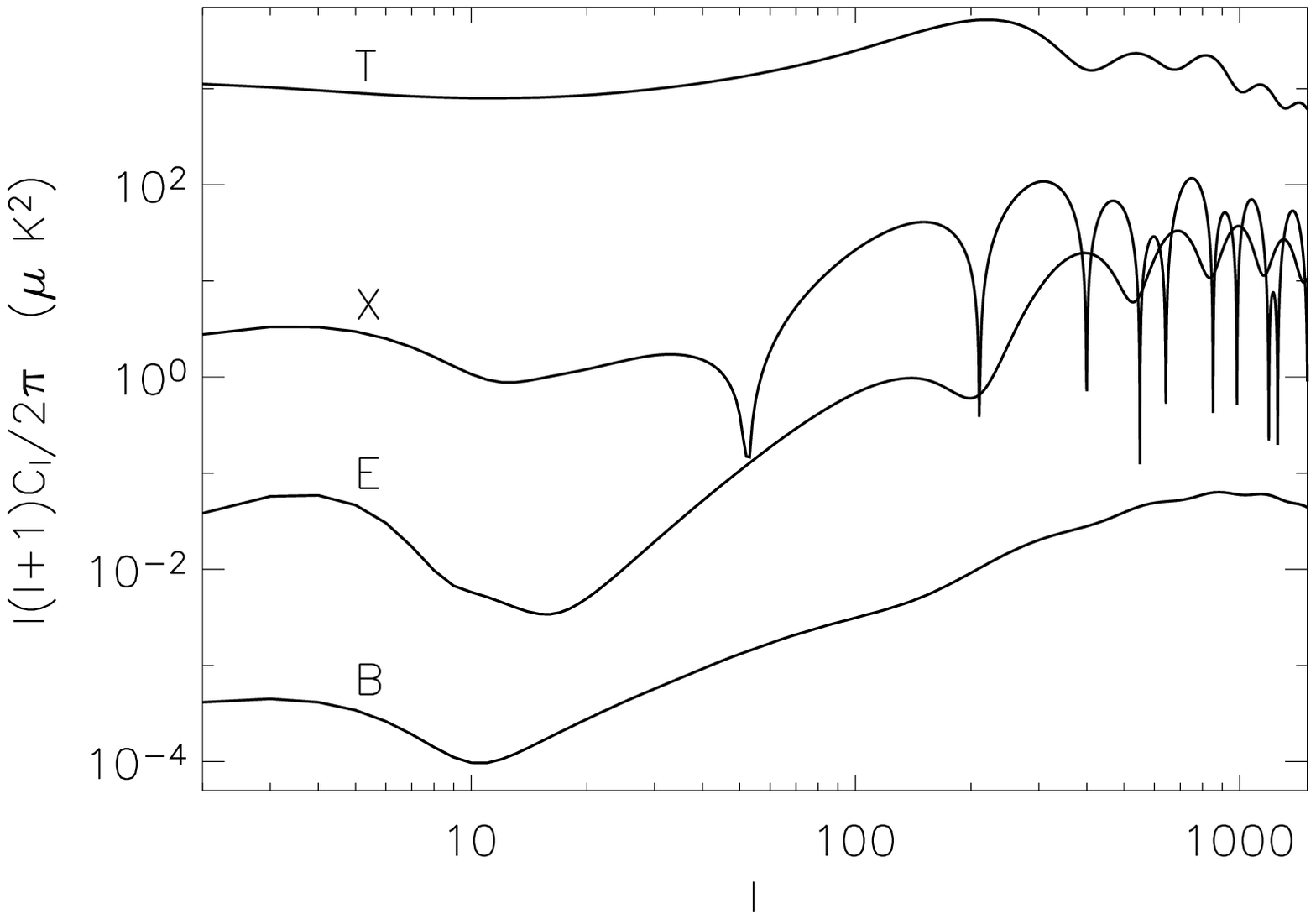}
\includegraphics[width=3in]{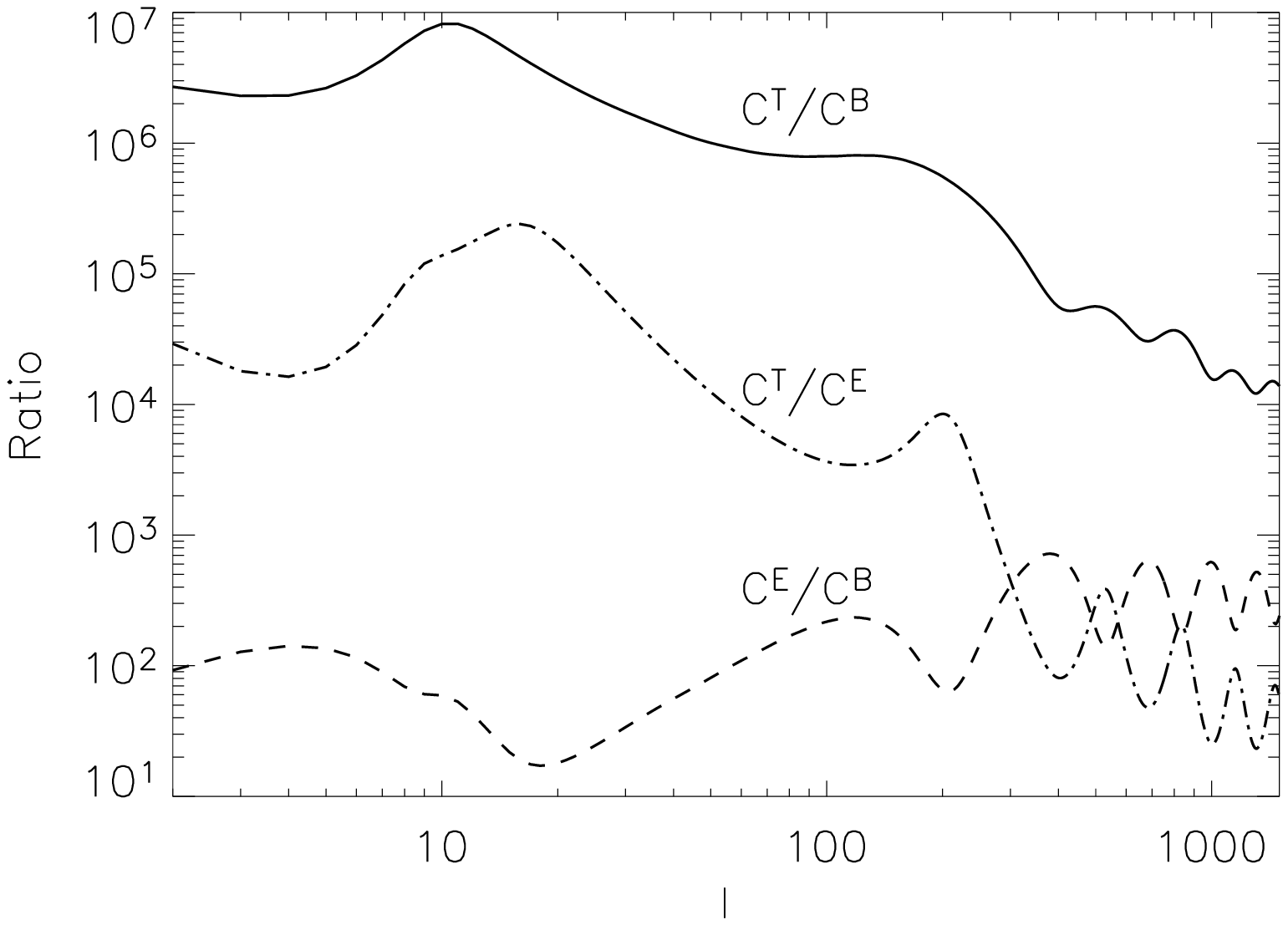}
\caption{Power spectra for temperature anisotropy (T),
$TE$ cross-correlation (X, absolute value plotted), $E$-type polarization, and $B$-type polarization.
The best-fit parameters from the three-year WMAP data were used \cite{wmap3imp}
with a tensor-to-scalar ratio $T/S=0.01$.  The right panel shows the ratios
of the power spectra.}
\label{fig:spectra}
\end{figure*}

A crucial insight into the analysis of CMB polarization 
data is the fact that 
any CMB polarization map can be divided into two components, a scalar
component, traditionally denoted $E$, and a pseudoscalar component
called $B$.  The CMB is weakly polarized, meaning that both of these
components are much smaller than the unpolarized (temperature)
anisotropy.  
Furthermore, the $B$ component is expected to be much
weaker than $E$, since scalar density perturbations produce only $E$
to linear order \cite{zalsel,selzal,kkslett,kks}.  (See Figure \ref{fig:spectra}.)
Experiments to date
have detected only the $E$ component.  In the future, the search for
the weaker $B$-type polarization will be a high priority, as the $B$
modes may contain the imprint of gravitational waves produced during
inflation.

Characterization of CMB polarization requires both very low
noise and exquisite control of systematic errors.  
In particular, 
some sources of systematic error may cause the polarization signal
to be contaminated by the much larger unpolarized anisotropy,
while others mix the $E$ and $B$ components.
As efforts to design $B$-mode experiments intensify, it is important
to consider carefully the susceptibility of different designs
to various kinds of error.  Hu \etal~\cite{HHZ} 
have provided a detailed framework for
performing such an analysis in the context of an imaging experiment.
For interferometric measurements, the issues are somewhat
different.
The purpose of this paper is to forecast the
effects of a variety of systematic errors on interferometric
measurements.

Interferometric methods have played an important role
in measurements of CMB anisotropy and polarization.
Pioneering attempts to detect CMB anisotropy with interferometers
are described in \cite{MarPar} and \cite{Sub}.
Several groups have successfully detected primary CMB anisotropies 
\citep{CAT1,CAT2,DASIT,CBIT,VSA}
and polarization
\citep{readhead,leitch}
using interferometers.  The formalism for analyzing CMB data from
interferometers has been developed by
a number of authors
\cite{HobLasJon,HobMag,WCDH,HobMais,Mye,BW}
as well as in the experimental papers cited above.

In any data set that fails to cover the entire sky, it is impossible
to separate the $E$ and $B$ components perfectly 
\cite{LCT,bunn,bunnerratum,bunnetal}.  The operation
of separating a polarization map into $E$ and $B$ components is nonlocal
when the map is viewed in real space, but in Fourier space or
spherical harmonic space, it can be done locally (mode by mode).
Since interferometric data sample the sky in the Fourier domain,
$E$-$B$ separation may be cleaner for interferometric
data than for maps made with single-dish instruments \cite{Parketal,ParkNg}.

As we will see, a variety of systematic errors in interferometers
can be modeled via Jones matrices \cite{tinbergen,heiles,HHZ} and by
deviations of the antenna patterns (including cross-polar contributions)
from assumed ideal forms.  We will assume that each of these errors
can be characterized by small unknown parameters, such as gain fluctuations,
cross-talk between detectors, pointing errors, etc.  We will
first calculate the
effect of each error on the measured visibilities.  We will
then provide a method
of quantifying the effects of each of these errors on estimates
of the polarization power spectra $C_l^E,C_l^B$ that can be obtained
from a hypothetical data set.  

This paper has the following structure.  Section \ref{sec:formalism} 
presents the mathematical
formalism we will use to describe interferometric visibilities for
polarization data.
Section \ref{sec:viserr} presents the effects of various systematic errors
on the visibilities extracted from a hypothetical CMB experiment.
Section \ref{sec:spectra} presents a method of forecasting errors on power spectrum
estimates from errors on visibilities.  Sections \ref{sec:instresults} 
and \ref{sec:beamresults} contain 
results showing how the error forecasts on both $E$ and $B$ 
power spectra depend on the parameters that characterize the various
systematic errors.  Section \ref{sec:discuss} presents a discussion of the
implications of these results, and
a brief appendix contains a useful mathematical result.

Sections \ref{sec:spectra} through \ref{sec:beamresults} 
contain quite a bit of technical detail.
The particularly busy or impatient reader should note that the
key ideas of Section \ref{sec:spectra} are summarized at the beginning, and the
final results of Sections \ref{sec:instresults} and \ref{sec:beamresults} 
are summarized in Section \ref{sec:discuss}
and Table \ref{table}.

\section{Formalism}
\label{sec:formalism}

\subsection{Antenna Patterns}
Consider a monochromatic plane wave of angular frequency 
$\omega=2\pi c/\lambda$
approaching the origin from a direction
$\rhat$.  The electric field of the wave is
\beq
{\bf E}({\bf x},t; \rhat)=\bein(\rhat) 
e^{i({\bf k}\cdot{\bf x}-\omega t)},
\eeq
where the wave vector ${\bf k}=-(\omega/c)\rhat$.  
(Of course,
as usual the physical electric field is just the real part of this
complex quantity.)

The complex vector 
$\bein(\rhat)$
is perpendicular to $\rhat$.  Its direction fluctuates rapidly
in time (except in the case of completely polarized radiation).  As usual,
all observables will be averages taken over a time that is long
compared to those fluctuations.  To be specific,
let $(\ein_X,\ein_Y)$ be Cartesian components of $\bein$ in the
plane perpendicular to $\rhat$, and 
define a  $2\times 2$ matrix ${\bf S}$ with
components
\beq
S_{ij}\propto\langle E_iE_j^*\rangle=\langle\ein_i\einsup{*}_j\rangle,
\label{eq:sij}
\eeq
where $i,j$ range over $X,Y$.  The matrix
${\bf S}$ can be expressed in terms of the standard Stokes parameters
\beq
{\bf S}=\begin{pmatrix}I+Q & U+iV\\ U-iV & I-Q\end{pmatrix}.
\label{eq:stokeslin}
\eeq
The constant of proportionality in equation (\ref{eq:sij}) depends on the
system of units being used.  We will follow the common practice in CMB
studies of expressing all Stokes parameters in dimensionless ``$\Delta T/T$''
form.

In general the electric
field at any given point will be a superposition of waves coming
in from all directions $\rhat$.  
We assume that the temporal fluctuations 
in the incoming waves from two distinct directions are uncorrelated
with each other:
\beq
\langle\ein_i(\rhat_1)\ein_j(\rhat_2)^*\rangle=
S_{ij}(\rhat_1)\delta(\rhat_1-\rhat_2),
\label{eq:eincorr}
\eeq
with $S_{ij}$ related to the Stokes parameters as above.

Consider a single-mode antenna that is designed to be sensitive to only one 
polarization direction.  We can model this
antenna as a device that sums all of the incoming
radiation, weighted by some vector-valued 
antenna pattern ${\bf a}$, to produce an output
that is the real part of
\beq
\eout=\int d^2\rhat\, {\bf a}(\rhat)\cdot\bein(\rhat)
e^{i({\bf k}\cdot{\bm\xi}-\omega t)},
\eeq
where ${\bm\xi}$ is the location of the antenna.

Throughout this paper, we will consider experiments in which the beam
width is small enough that the flat-sky approximation is
appropriate in analyzing any single pointing of the instrument.  (Mosaicking
of multiple pointings of such an instrument is considered in \cite{BW}.)
In
that case, we can represent the direction $\rhat$ by a vector in the
plane (specifically, the tangent plane to the sphere at the pointing
center) with components $(x,y)$.  In this approximation, for instance,
an antenna with a Gaussian beam pattern that is sensitive only to
linear polarization the $\hat{\bf x}$ direction would have ${\bf
a}\propto \hat{\bf x}\exp[-(x^2+y^2)/(4\sigma^2)]$, while an antenna
that is sensitive to either right or left circular polarization would
have $\hat{\bf x}\pm i\hat{\bf y}$ in place of $\hat{\bf x}$.  The
factor 4 in the Gaussian is present because we would like $\sigma$ to
be the Gaussian width of $|{\bf a}|^2$, not ${\bf a}$.

In some experiments, the same antenna may be used to measure two polarization
states (typically either ``horizontal'' and ``vertical'' 
linear polarization, or left
and right circular polarization).  In that case, the output of the
antenna will be a two-component vector $\beout$, and the antenna pattern
will be a $2\times 2$ matrix ${\bf A}$:
\beq
\beout=\int d^2\rhat\, {\bf A}(\rhat)\cdot\bein(\rhat)
e^{i({\bf k}\cdot{\bm\xi}-\omega t)}.
\label{eq:antint}
\eeq
In general, we will model antennas as sensitive to two polarization
states in this manner.  If in a particular experiment only one output
is actually
measured, we can simply ignore the other component of $\beout$.

We can of course express the components of the vectors $\bein$ and $\beout$
in any basis we like.  In particular, we can
resolve these vectors in either a linear polarization basis with
components $(\epsilon_X,\epsilon_Y)$ or a right- and left-circular
basis with components $(\epsilon_R,\epsilon_L)$.
The two bases
are related
by a unitary transformation
\beq
\begin{pmatrix}\epsilon_R\\ \epsilon_L\end{pmatrix}={\bf R}_{\rm circ}\cdot
\begin{pmatrix}\epsilon_x\\ \epsilon_y\end{pmatrix},
\eeq
with
\beq
{\bf R}_{\rm circ}=
\frac{1}{\sqrt{2}}\begin{pmatrix}
1 & i\\
1 & -i 
\end{pmatrix}.
\label{eq:rcirc}
\eeq
In either case, an ideal antenna, i.e., one with equal response 
to both polarization states and no mixing between them, would have 
${\bf A}$ equal to a scalar function times the identity matrix.

In the circular polarization basis, the elements of the Stokes parameter
matrix ${\bf S}$ become
\beq
{\bf S}_{\rm circ}={\bf R}_{\rm circ}\cdot{\bf S}_{\rm lin}\cdot
{\bf R}_{\rm circ}^{-1}
\eeq
with ${\bf S}_{\rm lin}$ given by equation (\ref{eq:stokeslin}).
Explicitly, we have
\beq
{\bf S}_{\rm circ}=\begin{pmatrix} I+V & Q+iU \\ Q -iU & I-V\end{pmatrix}.
\label{eq:stokescirc}
\eeq

\subsection{Visibilities}
\label{sec:vis}
Consider an interferometer with $N$ antennas.  The output signal from
antenna $j$ will be denoted $\beoutsup{(j)}$.  As noted earlier,
we will treat this as a two-component vector with components
$\eoutsup{(j)}_m$, with $m=X,Y$ for a linear polarization experiment
or $m=R,L$ for a circular polarization experiment.  

The basic datum
for an interferometer is a ``visibility'' obtained by correlating a component
of $\eout$ from one antenna with a component from another antenna:
\beq
V_{mn}^{(jk)}=\langle \eoutsup{(j)}_m\eoutsup{(k)*}_{\!\! n}\rangle,
\eeq
where the angle brackets denote a time average.
Both real and imaginary parts of this complex quantity
can be obtained by measuring the in-phase and quadrature-phase
correlations.

For a fixed pair of antennas $(jk)$, the visibilities form a $2\times 2$
matrix ${\bf V}^{(jk)}$.
Using equations (\ref{eq:eincorr}) and (\ref{eq:antint}), we can write this 
matrix as
\beq
{\bf V}^{(jk)}=\int d^2\rhat\,
{\bf A}^{(j)}(\rhat)\cdot {\bf S}(\rhat)
\cdot {\bf A}^{(k)\dag}(\rhat)\ e^{-2\pi i {\bf u}_{jk}\cdot\rhat}.
\label{eq:stokestovis}
\eeq
Here the ${\bf A}$'s are the antenna patterns for the two antennas
and ${\bf u}_{jk}=(\bm{\xi}^{(k)}-\bm{\xi}^{(j)})/\lambda$ 
is the separation between the two antennas in units of wavelength.
The matrix ${\bf A}^\dag$ is the hermitian conjugate of ${\bf A}$
(that is, the complex conjugate of the transpose of ${\bf A}$).

In an ideal experiment with ${\bf A}$ proportional to the identity
matrix (no cross-polar response and identical co-polar response to
both polarization states), ${\bf V}^{(jk)}\propto{\bf S}$.  In other words,
each visibility measures a simple linear
combination of the Stokes parameters.  To be explicit, let us define
Stokes visibilities
\beq
V_{Z}^{(jk)}\equiv \int d^2\rhat\,A^{(j)}(\rhat)Z(\rhat)A^{(k)}(\rhat)
e^{-2\pi i{\bf u}_{jk}\cdot\rhat},
\eeq
where $Z=I,Q,U,V$ is a Stokes parameter.
As is well known, these can also be written as a convolution in Fourier 
space:
\beq
V_Z^{(jk)}\propto \int d^2{\bf k}\,\tilde{Z}({\bf k})
\widetilde{\cal A}_{jk}^*({\bf k}-2\pi{\bf u})
\label{eq:visfourier}
\eeq
with ${\cal A}_{jk}=A^{(j)}A^{(k)}$.

A polarimetric interferometer can work either by interfering
linear polarization states or circular polarization states.  Throughout
this paper, we
will refer to these possibilities as {\it linear experiments} and {\it
circular experiments} respectively.  
Information about
both linear and circular polarization can be obtained from either
type of experiment.

In an ideal linear
experiment, 
we would extract the Stokes parameters from the visibility
matrix as follows:
\begin{subequations}
\label{eq:stokesvislin}
\begin{align}
V_I&=\tfrac{1}{2}(V_{XX}+V_{YY}),\\
V_Q&=\tfrac{1}{2}(V_{XX}-V_{YY}),\label{eq:vqbad}\\
V_U&=\tfrac{1}{2}(V_{XY}+V_{YX}),\label{eq:visulin}\\
V_V&=\tfrac{1}{2i}(V_{XY}-V_{YX}).
\end{align}
\end{subequations}
Here we are assuming that all antennas split up the incoming radiation
into orthogonal linear polarizations with respect to a single 
fixed coordinate system $(X,Y)$.  The superscript $(jk)$ is suppressed.

For the weak polarization found in CMB data,
equation (\ref{eq:vqbad}) is not a practical way to measure Stokes $Q$
because it requires perfect cancellation of the much larger $I$ contributions;
in practice, such an experiment measures linear polarization only via
$U$, not $Q$.
Since $Q\to U$ under a $45^\circ$ rotation, we measure $Q$ in practice
by using antennas that measure linear polarization states 
in a basis $(X',Y')$ that is rotated with
respect to $(X,Y)$.  This can be done either by rotating the 
instrument or by having the polarizers on different antennas
oriented in different ways.  In either case, note that in general
the Stokes parameters $Q,U$ are not generally measured with the same
baseline at the same time.

The corresponding relations in a circular experiment are
\begin{subequations}
\label{eq:stokesviscirc}
\begin{align}
V_I&=\tfrac{1}{2}(V_{RR}+V_{LL}),\\
V_Q&=\tfrac{1}{2}(V_{RL}+V_{LR}),\label{eq:vqcirc}\\
V_U&=\tfrac{1}{2i}(V_{RL}-V_{LR}),\label{eq:vucirc}\\
V_V&=\tfrac{1}{2}(V_{RR}-V_{LL}).
\end{align}
\end{subequations}
In a circular experiment, both $Q$ and $U$ visibilities can be measured
simultaneously on a single baseline.

We do not expect any cosmological source of circular polarization:
Stokes $V$ is expected to be zero.  Nonetheless,
it may be useful to measure the Stokes visibility $V_V$ as a monitor
of systematic errors.  Conversely, if a non-cosmological source of circular
polarization is present, systematic errors may cause it to contribute
to measurements of linear polarization.

\subsection{Modeling systematic errors}

A wide variety of systematic errors can be modeled as imperfections
in the matrix-valued antenna patterns of the antennas.   We will model
these errors in the following way:
\beq
{\bf A}(\rhat) = {\bf J}_i\cdot{\bf R}\cdot{\bf A}_s(\rhat)
\cdot {\bf R}^{-1}.
\label{eq:antennajones}
\eeq
Here ${\bf R}$ is ${\bf R}_{\rm circ}$ for a circular
experiment and is the identity matrix for a linear experiment.

The ``instrument Jones matrix'' ${\bf J}_i$ represents errors introduced within
the instrument, such as gain errors and cross-talk between the two
outputs of a given antenna. The matrix ${\bf A}_s$ is the antenna
pattern on the sky, before such instrumental errors are taken into
account.  We use ${\bf A}_s$ to model cross-polarization, beam errors,
pointing errors, etc.
We will always use ${\bf A}$ (with no subscript) to denote
the overall antenna pattern including both ``instrument'' and ``sky''
effects.

We will always represent ${\bf A}_s$ in a Cartesian basis;
that is, it acts on components $(\ein_X,\ein_Y)$.
When we are performing a circular
experiment, however, $\bein$ and $\beout$ will be represented
in a circular polarization basis.  The factors ${\bf R}$ 
and ${\bf R}^{-1}$
are inserted to account for this change of basis.

An ideal instrument would have ${\bf J}_i={\bf 1}$, the identity
matrix, and ${\bf A}_s=A(\theta,\phi){\bf 1}$.

There is some redundancy in equation (\ref{eq:antennajones}).
Mathematically, we could correctly describe any instrument without
including ${\bf J}_i$ by simply absorbing its effects into ${\bf
A}_s$.  However, it is convenient to maintain the distinction
between effects that happen to the signal before the antenna averages
over the beam (effects ``on the sky'')
and afterwards (effects ``in the instrument'').  Instrument errors
are easier to model, because by definition they do not depend on 
position on the sky.

\section{Effect of errors on visibilities}
\label{sec:viserr}

In this section we compute the effects of various sorts of instrument
and beam errors on the measured visibilities $V_Q,V_U$.   Each
of the errors considered can be modeled with a
set of small parameters.  We assume that the experimenter
has no knowledge of these errors (or else she would have removed them)
and hence analyzes the data under the assumption that the experiment
is error-free.  

\subsection{Instrument Errors}
\label{sec:inst}
Consider first the effect of errors within the instrument, assuming
for the moment that ${\bf A}_s$ is of the ideal form $A_s(\rhat){\bf 1}$.
We can completely characterize the instrumental Jones matrix for the $j$th
antenna with gain
errors $g_1^{(j)},g_2^{(j)}$ and couplings $\epsilon_1^{(j)},\epsilon_2^{(j)}$:
\beq
{\bf J}_i^{(j)}=
\begin{pmatrix}
1+g_1^{(j)} & \epsilon_1^{(j)} \\
\epsilon_2^{(j)} & 1+g_2^{(j)}
\end{pmatrix}.
\label{eq:jones}
\eeq
This is similar to the characterization in ref.~\cite{HHZ}, although
our notation is not identical to theirs.  In particular, 
we treat the $g$ and $\epsilon$
parameters as complex numbers (to account for arbitrary phases in the
errors) rather than introducing explicit phase angles.
The parameters $g$ and
$\epsilon$ will be assumed to be small (i.e., products of them
will be neglected).

A number of different physical effects can be encoded in
a matrix of this form.  The gain parameters $g_i^{(j)}$ incorporate
both errors in the magnitude of the gain and unaccounted-for phase
delays.  The couplings $\epsilon_i^{(j)}$ can account for mixing
of the two polarization states within the optical and electronic systems
and also,
in the case of a linear experiment, for an error in alignment
of the polarizers: if the polarizers in antenna $j$ are misaligned
by an angle $\delta$, then
$\epsilon_1^{(j)}=\delta$,
$\epsilon_2^{(j)}=-\delta$.

By using this model for the antenna patterns in equation 
(\ref{eq:stokestovis}), we can determine the effect of all of these errors
on the recovered Stokes visibilities.  The results depend on whether
we are considering a linear or circular experiment.

\medskip
\textit{Linear experiment:}
Information about linear polarization in such an
experiment comes from the visibility for Stokes $U$.
Using equations (\ref{eq:stokeslin}),
(\ref{eq:stokestovis}), and (\ref{eq:visulin}), we find
\begin{align}
V_U^{(jk)}=\Vo_U^{(jk)} & +
\frac{1}{2}  \left[
\Vo_I^{(jk)}(\epsilon_1^{(j)}+\epsilon_2^{(j)}+\epsilon_1^{(k)*}
+
\epsilon_2^{(k)*})
\right.\nonumber\\
&
+\Vo_U^{(jk)}(g_1^{(j)}+g_2^{(j)}+g_1^{(k)*}+g_2^{(k)*})\nonumber\\
&+ \Vo_Q(-\epsilon_1^{(j)}+\epsilon_2^{(j)}-\epsilon_1^{(k)*}+
\epsilon_2^{(k)*})\nonumber\\
&+ \left.\Vo_V(g_1^{(j)}-g_2^{(j)}-g_1^{(k)*}+g_2^{(k)*})\right],
\label{eq:instlin}
\end{align}
working to linear order in the small quantities.
Here the symbol $\Vo$ indicates the visibility that would
be measured in the absence of systematic errors.  Note
that the coupling parameters $\epsilon$
mix temperature anisotropy ($I$) into polarization; this
is in general the most serious sort of error.  The
$\Vo_Q$ and $\Vo_U$ 
terms are less worrisome, since they only involve polarization.
However, as we will see they do couple $E$ to $B$ and so
can be serious for a $B$-mode experiment.

Although for cosmological purposes
we are primarily interested in measurements of linear polarization, 
with this experimental setup we get the visibility for Stokes $V$ ``for free''
by subtracting rather than adding $V_{XY}$ and $V_{YX}$ [see equations
(\ref{eq:stokesvislin})].  Assuming there is no intrinsic circular
polarization ($\Vo_V=0$), the leading contribution is
\beq
V_V^{(jk)}=\frac{i}{2}\Vo_I^{(jk)}(\epsilon_1^{(j)}-\epsilon_2^{(j)}
-\epsilon_1^{(k)*}+\epsilon_2^{(k)*}),
\eeq
neglecting terms proportional to $\Vo_Q,\Vo_U$.  Even very low
levels of coupling may therefore provide a measurable signal, which would
be correlated in a known way with the temperature map.  This may provide
a useful diagnostic.

Conversely, if there is intrinsic circular polarization (e.g.,
due to foregrounds, equation (\ref{eq:instlin}) shows that gain
errors can couple that signal into $V_Q,V_U$.

\medskip\textit{Circular experiment:}
Now suppose we perform an experiment in which interference 
between right and left circular
polarization states is measured.
Using equations (\ref{eq:stokescirc}),
(\ref{eq:stokestovis}), (\ref{eq:vqcirc}) and (\ref{eq:vucirc}), we find
\begin{subequations}
\label{eq:instcirc}
\begin{align}
V_Q^{(jk)}
=&\Vo_Q^{(jk)}+\frac{1}{2}\left[
\Vo_I^{(jk)}(\epsilon_1^{(j)}+\epsilon_2^{(j)}+\epsilon_1^{(k)*}+
\epsilon_2^{(k)*})\right.\nonumber\\
&+
\Vo_Q^{(jk)}(g_1^{(j)}+g_2^{(j)}+g_1^{(k)*}+g_2^{(k)*})\nonumber\\
& +i\Vo_U^{(jk)}(g_1^{(j)}-g_2^{(j)}-g_1^{(k)*}+g_2^{(k)*})\nonumber\\
& +\left.
\Vo_V^{(jk)}(-\epsilon_1^{(j)}+\epsilon_2^{(j)}-\epsilon_1^{(k)*}+
\epsilon_2^{(k)*})\right],\label{eq:instcirc1}\\
V_U^{(jk)}
=&\Vo_U^{(jk)}+\frac{1}{2}\left[
i\Vo_I^{(jk)}(-\epsilon_1^{(j)}+\epsilon_2^{(j)}+\epsilon_1^{(k)*}-
\epsilon_2^{(k)*})\right.\nonumber\\
& +
\Vo_U^{(jk)}(g_1^{(j)}+g_2^{(j)}+g_1^{(k)*}+g_2^{(k)*})\nonumber\\
& -i\Vo_Q^{(jk)}(g_1^{(j)}-g_2^{(j)}-g_1^{(k)*}+g_2^{(k)*})\nonumber\\
& +\left.
i\Vo_V^{(jk)}(\epsilon_1^{(j)}+\epsilon_2^{(j)}-\epsilon_1^{(k)*}-
\epsilon_2^{(k)*})\right].\label{eq:instcirc2}
\end{align}
\end{subequations}
As in the linear case, coupling errors ($\epsilon$) are 
the main danger, causing leakage
from $I$ into $Q,U$.  
Gain errors mix $V_Q$ and $V_U$ with each other.  As we will see,
this means that their effect on the $B$ mode power spectrum is somewhat
more severe than in the case of a linear experiment.

As for linear experiments, we might consider monitoring the
circular polarization visibility $V_V$ as a diagnostic,
even though we do not expect any cosmological signal.
In this case, the leading term in $V_V$ is proportional
to the gain fluctuations $g_i^{(j)}$ and to $\Vo_I$.
Since gain errors are less worrisome than couplings, this
may not be as valuable as in the linear case.  Furthermore,
unlike a linear
experiment, in a circular experiment one does not necessarily get
$V_V$ for free when measuring $V_Q,V_U$: the linear polarization
information is obtained by interfering right with left polarization
states,
while $V_V$ comes from interfering identical ones [see equation 
(\ref{eq:stokesviscirc})].

\subsection{Beam Errors}
\label{sec:beam}

We next consider errors that can be modeled via the ``sky'' 
matrix ${\bf A}_s$.  In this section we ignore instrument errors, taking 
${\bf J}_i$ to be the identity matrix.  

An ideal experiment, with ${\bf A}_s$ equal to a scalar function times
the identity matrix, would have identical response to both
polarization states and no mixing between them.
Furthermore, ideally ${\bf A}_s$ would be the same for all antennas.
There are of course a large number of ways these idealizations
can fail.  Unlike instrument errors, which are characterized by a finite
list of parameters, beam errors are characterized by arbitrary
{\it functions} on the sky.  Rather than providing
a complete catalogue of this infinite space of possibilities, we 
focus on a few physically motivated possibilities.

\medskip\textit{Beam mismatch:}
We first consider the case where each antenna pattern is
proportional to the identity matrix, but the various
antenna patterns differ from each other and from the form
assumed by the experimenter:
\beq
{\bf A}_s^{(j)}(\rhat)=
A^{(j)}(\rhat){\bf 1}.
\label{eq:mismatch}
\eeq
In this case, we are assuming no cross-polar response and identical
beam patterns for both polarizations in each antenna.
This formulation can account for
pointing errors as well as errors in beam shape (e.g., beam width
and ellipticity errors).


Assuming this form for the antenna pattern, we can
use equation (\ref{eq:stokestovis})
to extract the Stokes visibilities, yielding
\begin{equation}
V_{Z}^{(jk)}
=\int d^2\rhat\,e^{-2\pi i{\bf u}_{jk}\cdot\rhat}
Z(\rhat)A^{(j)}(\rhat)
A^{(k)*}(\rhat).
\label{eq:beamvis}
\end{equation}
For $Z=\{I,Q,U,V\}$.  These results apply to both linear and circular experiments,
although in practice only $V_U$ would be used for linear
polarization information in a linear experiment.

This category of error causes no leakage from $I$ into polarization
or even between $Q$ and $U$.  Nonetheless, as we will see it can cause
$E/B$ mixing when the power spectra are estimated.

\medskip\textit{Cross-polarization:}
We now consider the possibility of cross-polar antenna response (i.e.,
off-diagonal entries in ${\bf A}_s$).  For simplicity, we consider
only the case of an azimuthally symmetric antenna.  
We therefore begin by determining what form of ${\bf A}_s$
are compatible with azimuthal symmetry.

First, note that ${\bf A}_s$ must be diagonal when measured along
the $x$ axis of our coordinate system.  One way to see this is to
invoke reflection symmetry: if $\ein_x$ coupled to $\eout_y$
along the $x$ axis, the coupling would have to change sign upon reflection,
but a symmetric antenna should be invariant under reflections.
Using polar
coordinates $(r,\phi)$, we can therefore write
\beq
{\bf A}^{(i)}_s(r,0)=\begin{pmatrix}A^{(i)}_0(r)+\frac{1}{2}A^{(i)}_1(r) & 0 \\ 
0 & A^{(i)}_0(r)-\frac{1}{2}A^{(i)}_1(r)
\end{pmatrix}
\label{eq:xpdiag}
\eeq
for arbitrary radial functions $A^{(i)}_0,A^{(i)}_1$.
Suppose we now measure the antenna pattern for some nonzero $\phi$.  We
must get the same answer as at $\phi=0$, as long as we transform both $\bein$
and $\beout$ by a rotating through an angle $-\phi$ to move them to the $x$ axis.  
We conclude that 
${\bf A}_s(r,\phi)={\bf R}_{-\phi}\cdot {\bf A}_s(r,0)\cdot{\bf R}_{\phi}$,
where ${\bf R}_{\pm\phi}$ are the appropriate rotation matrices.  Performing
the matrix multiplication, we find that
\beq
{\bf A}_s^{(i)}=\begin{pmatrix}A_0^{(i)}+\frac{1}{2}A_1^{(i)}\cos 2\phi &
\frac{1}{2}A_1^{(i)}\sin 2\phi \\
\frac{1}{2}A_1^{(i)}\sin 2\phi & A_0^{(i)}-\frac{1}{2}A_1^{(i)}\cos 2\phi
\end{pmatrix}.
\label{eq:xpantenna}
\eeq


Assuming the form (\ref{eq:xpantenna}) for ${\bf A}_s$, we obtain the following
expressions for the visibilities:
\begin{subequations}
\label{eq:xpvis}
\label{eq:crossvis}
\begin{align}
V_Q^{(jk)}&=
\int d^2\rhat\,e^{-2\pi i{\bf u}_{jk}\cdot{\bf r}}
\left\{
Q(\rhat)A_0^{(j)}(r)A_0^{(k)*}(r)+\right.\nonumber\\
&.\tfrac{1}{2}I(\rhat)[A_0^{(j)}(r)A_1^{(k)*}(r)+A_1^{(j)}(r)A_0^{(k)*}(r)]\cos 2\phi+\nonumber\\
&\left.
\tfrac{i}{2}V(\rhat)[A_0^{(j)}(r)A_1^{(k)*}(r)-A_1^{(j)}(r)A_0^{(k)*}(r)]
\sin 2\phi
\right\},\\
V_U^{(jk)}&=
\int d^2\rhat\,e^{-2\pi i{\bf u}_{jk}\cdot{\bf r}}
\left\{
U(\rhat)A_0^{(j)}(r)A_0^{(k)*}(r)+\right.\nonumber\\
&\tfrac{1}{2}
I(\rhat)[A_0^{(j)}(r)A_1^{(k)*}(r)+A_1^{(j)}(r)A_0^{(k)*}(r)]\sin 2\phi-
\nonumber\\
&\left.
\tfrac{i}{2}V(\rhat)[A_0^{(j)}(r)A_1^{(k)*}(r)-A_1^{(j)}(r)A_0^{(k)*}(r)]
\cos 2\phi
\right\},
\end{align}
\end{subequations}
neglecting terms that are quadratic in the small quantities $A_1$.
Again, these results apply to both linear and circular experiments, although
only $V_U$ is used for polarization measurements in a linear
experiment.  We immediately see that cross-polarization has the danger of
coupling $I$ to polarization.

If we write down a similar expression for the circular polarization
visibility $V_V$, we find terms proportional to Stokes $Q$ and $U$
but not $I$;  thus no great insight into cross-polarization is likely to
be found by monitoring $V_V$.  On the other hand, if there is a
strong intrinsic
circular polarization signal from foregrounds, cross-polarization may
cause it to contaminate the linear polarization observables.

\section{Effects on Power Spectra}
\label{sec:spectra}

\subsection{Introduction}
\label{sec:specintro}

The primary goal of almost any CMB experiment is to measure
some or all of the temperature and polarization power spectra.
We must therefore consider how to propagate the errors described
above to obtain forecasts of the effects of various systematic
errors on power spectrum estimates.
The completely correct approach
to this question is to define a precise experimental setup and
simulate it in detail.  We will instead adopt an approach that is 
simpler and more general.  

The power spectra are well-localized in Fourier or spherical
harmonic space and therefore not at all localized in image space.
Interferometric visibilities are fairly well-localized in Fourier
space, so power spectrum estimates can be obtained from individual
visibilities (or in the case of polarization, from pairs of
visibilities $V_Q,V_U$).  This is in contrast to an imaging
experiment, in which the entire map goes into each power spectrum
estimate.  We can therefore assess the effects of various errors on
power spectrum estimates by working with one visibility pair at a
time.  Naturally, analysis of real data would be considerably more
sophisticated than the process we consider here, but this approach
is adequate for assessing the rough scale at which each systematic
error affects power spectrum estimates.  The rest of this section
will present this approach in detail.

A data set from an interferometric experiment consists of measurements
of the visibilities $V_Q,V_U$ for many different baseline separations
and instrument orientations -- that is, for many different values of ${\bf u}$.
Let us assume that our data set consists of a set of visibilities
$V_{Q1},V_{U1},V_{Q2},V_{U2},\ldots,V_{QN},V_{UN}$.  We will assume
that $V_{Qi},V_{Ui}$ are both measured with the same baseline vector
${\bf u}_i$, and furthermore that distinct baselines ${\bf u}_i,{\bf
u}_j$ are far enough apart in the Fourier plane that we can treat the
corresponding visibilities as uncorrelated: $|{\bf u}_i-{\bf
u}_j|\gtrsim \Delta u$, where the Fourier-space resolution $\Delta u$
scales inversely with the beam width.
In a real experiment, these assumptions would presumably
not be true: at least some regions of 
the visibility plane would be oversampled, and in the case
of a linear experiment the two Stokes parameters would not always
be measured with identical baselines.  In such a
case, we can imagine binning the visibilities so as to consider a
smaller number of effectively independent samples in the visibility
plane.

In this approximation, each set of visibilities $(V_{Qi},V_{Ui})$
gives us an independent estimate of the power spectra $C_l^E,C_l^B$ at
$l\approx 2\pi u_i$.  (Here ``independent'' means ``independent of
the estimates obtained from the other visibilities''; unfortunately,
it is not in general the case
that the estimates of $C_l^E$ and $C_l^B$ are independent of each
other.)  We therefore begin by assessing the effect of each systematic
error on the power spectrum estimates derived from a single pair of
visibilities $(V_Q,V_U)$ for some fixed baseline vector ${\bf u}$.  We
assume that systematic errors must be controlled at least well enough
that the fractional errors introduced in the power spectrum 
estimates from each individual visibility are small.

In this section, we begin with the optimal
estimators of $E$ and $B$ band powers from a single visibility
pair $(V_Q,V_U)$ under the idealized assumption that there are no
errors in the experiment.  We then imagine ``turning on'' sources
of systematic error, one at a time, and show how to calculate the 
resulting root-mean-square (r.m.s.) band power errors.  
We will find that
the r.m.s.\ error
induced by any particular systematic error can be written in the form
\beq
(\delta\hat C_{\rm rms}^{K})^2=p_{\rm rms}^2\sum_{I,J}\kappasubsq{K,IJ,p}C^IC^J.
\label{eq:mainresult}
\eeq
Here $p$ is a parameter characterizing the strength of the
source of error, (e.g., one of the gain fluctuation parameters $g_i^{(j)}$).
The superscript $K=\{E,B\}$ indicates the type of power spectrum
being measured, and $\delta\hat C_{\rm rms}^K$ is the error
in an estimate of a band power.
The quantities $C^{I},C^J$ are band powers 
with
$I,J$ ranging over $\{T,X,E,B\}$ (temperature, $TE$ cross-correlation, and
$E$ and $B$ polarization spectra).
We will see how to calculate the coefficients $\kappa$ below.

As Figure \ref{fig:spectra} shows, there is 
a clear hierarchy in the input power spectra:
$C^T>C^X>C^E>C^B$.  The above sum is therefore usually dominated
by one or at most two terms that couple large spectra to smaller ones.
For each of the sources of error
described in the previous sections, we can find the one or two most serious
such couplings and thus estimate the level to which the overall
error must be controlled or at least understood.

As noted above, our approach is to demand that errors be understood to a level
low enough that they do not contaminate band powers measured from
each individual baseline.
This may be regarded as a conservative approach: even if an error
has quite a large effect on power spectrum estimates from a single
baseline, it may still be possible to extract good power spectrum
estimates by combining results from a set of many baselines.  
This is particularly true if the visibility plane is heavily oversampled, 
with many different visibilities within a resolution element $\Delta u$,
{\it if} the errors associated with the various visibilities
are independent and average to zero.
However, since systematic
errors are in general non-Gaussian and time-varying, one would
not want to rely on such an averaging procedure unless
the statistical properties of the errors are understood quite
well.  It therefore seems reasonable to adopt such a conservative approach.

\subsection{Ideal estimators}

We begin with the estimators for $E$ and $B$ band powers for
a single visibility pair $(V_Q,V_U)$, for an idealized experiment
with all systematic errors ``turned off.''  In the following
subsection, we will examine the effect on these estimators
when the errors are introduced.

Let ${\bf v}\equiv (V_I,V_Q,V_U)$ be a set of visibilities
corresponding to a single baseline ${\bf u}$.
Our hypothetical experiment measures $V_Q$ and $V_U$, but not
necessarily $V_I$.  As we will see, it is convenient
to include $V_I$ in the formalism anyway, because
it may be coupled to the others by various systematic
errors.

In the absence of systematic errors,
the $I$ visibility measures only the temperature fluctuations:
\beq
\langle |V_I|^2\rangle = C^T_{2\pi u}.
\eeq
Here $C_{2\pi u}^T$ is a band power giving 
the average temperature power spectrum in the vicinity of $l=2\pi u$.
We omit a constant of proportionality in this expression and
hereinafter by assuming 
that all visibilities have been scaled by an appropriate
factor.

\begin{figure}
\includegraphics[width=3in]{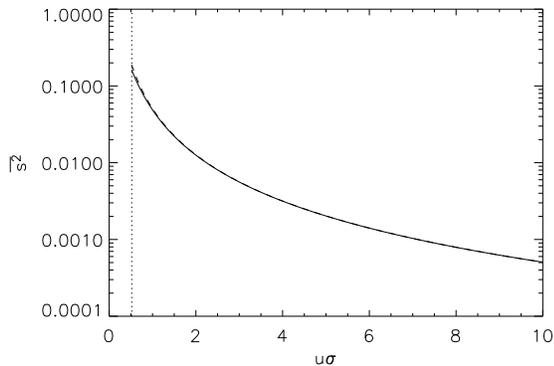}
\caption{The quantity $\overline{s^2}$ as a function of antenna separation.
The quantity on the horizontal axis is the product of baseline and
beam width and is proportional to the
separation between the antennas.  A Gaussian beam shape is assumed.
For antennas of diameter $D$ with FWHM$=1.22\lambda/D$, 
the vertical dotted line at $u\sigma=0.52$
corresponds to a pair of antennas that are just touching.
The dashed curve is the approximation (\ref{eq:ssqapprox}).}
\label{fig:sc}
\end{figure}

Suppose for the moment
that the baseline vector ${\bf u}$ points in the $x$ direction
of our chosen coordinate system.  The contributions
of $E$ and $B$ polarization to Stokes $Q$ and $U$ 
are expressed in Fourier space as follows:
\begin{subequations}
\begin{align}
\tilde Q&=\tilde E \cos(2\phi)-\tilde B\sin(2\phi),\\
\tilde U&=\tilde E \sin(2\phi)+\tilde B\cos(2\phi),
\end{align}
\end{subequations}
where $\phi$ is the angle made by the wave vector
${\bf k}$ with respect to the $x$
axis.  Each visibility is an average of $\tilde Q$ or $\tilde U$ over
a range of wave vectors centered on ${\bf k}=2\pi{\bf u}$ [see
eq.~(\ref{eq:visfourier})].  Assuming that the power spectra can be pulled
out of the integral to give band powers, we have
\begin{subequations}
\begin{align}
\langle |V_Q|^2\rangle &=C_{2\pi u}^E\csq +C^B_{2\pi u}\ssq,\\
\langle |V_U|^2\rangle &=C_{2\pi u}^E\ssq +C^B_{2\pi u}\csq,
\end{align}
\end{subequations}
where $\csq,\ssq$ are averages of $\cos^2 (2\phi),\sin^2(2\phi)$
over the antenna patterns:
\begin{subequations}
\label{eq:csqssq}
\begin{align}
\csq&=\frac{\int |\widetilde{A^2}
({\bf k}-2\pi {\bf u})|^2\cos^2(2\phi)\,d^2{\bf k}}
{\int |\widetilde{A^2}({\bf k}-2\pi {\bf u})|^2\,d^2{\bf k}},\\
\ssq&=\frac{\int |\widetilde{A^2}
({\bf k}-2\pi {\bf u})|^2\sin^2(2\phi)\,d^2{\bf k}}
{\int |\widetilde{A^2}({\bf k})|^2\,d^2{\bf k}}=1-\csq.
\end{align}
\end{subequations}
Here $\widetilde{A^2}$ is the Fourier transform of the squared antenna
pattern.
There can also be $T$-$E$ correlations, given by the cross-correlation
$C_l^X$.  These relate $V_Q$ to $V_I$:
\beq
\langle V_IV_Q^*\rangle=C_{2\pi u}^X\cc,
\eeq
where $\cc$ is an average of $\cos(2\phi)$ analogous to equations
(\ref{eq:csqssq}).
These expressions are valid only for
${\bf u}$
parallel to the $x$ axis.  Without this assumption, there would 
be a $V_I$-$V_U$ covariance proportional to $\overline s$, the
average of $\sin(2\phi)$, and a $V_Q$-$V_U$ covariance
proportional to $\overline{sc}$,
but with this geometry both of these terms vanish.

In the limit where the visibility is measured with two very widely
separated antennas ($2\pi u\sigma\gg 1$ for beam width $\sigma$), 
each visibility samples a very narrow
region in the Fourier plane.  In this case, 
$\csq\approx 1,\cc\approx 1$, and $\ssq\approx 0$.
The squares of
$V_I,V_Q,V_U$ then provide pure estimates of $C^T,C^E,C^B$
respectively.  

The quantity $\ssq$ characterizes 
mixing of $E$ and $B$ modes within each visibility pair.  As a result, the
degree to which systematic errors couple different power spectra
is strongly dependent on
$\ssq$.
Fig.~\ref{fig:sc} shows $\ssq$ as a function
of $u\sigma$, the product of baseline length and beam width.
This product is also proportional to the antenna separation in units
of antenna diameter.  In particular, for antennas with a Gaussian
beam pattern with ${\rm FWHM}=1.22\lambda/D$, the separation
is equal to the diameter (the antennas are just touching)
when $u\sigma=0.52$.  The quantity
$\ssq$ is well approximated by
\beq
\ssq=\frac{1}{2\pi^2(u\sigma)^2}.
\label{eq:ssqapprox}
\eeq

Given the visibilities $V_Q,V_U$, the optimal estimators
of the polarization band powers are
\begin{subequations}
\label{eq:ebest}
\begin{align}
\hat{C}^E&= \gamma(\csq |V_Q|^2-\ssq|V_U|^2),
\label{eq:eest}\\
\hat{C}^B&= \gamma(\csq |V_U|^2-\ssq|V_Q|^2),
\label{eq:best}
\end{align}
\end{subequations}
suppressing the subscripts $2\pi u$,
with
\beq
\gamma=[(\csq)^2-(\ssq)^2]^{-1}.
\eeq

One can check using the covariances
above that these expressions give unbiased estimates.
%
Furthermore, it is straightforward but tedious to check that $\hat C^E,
\hat C^B$
are the maximum-likelihood estimators in the case of Gaussian fluctuations.
The Cram\'er-Rao inequality (e.g., \cite{berger}) then implies
that they are the optimal estimators.

It will be convenient to write these estimators in fancier linear-algebra
language.  The covariances of the Stokes visibilities ${\bf v}=(V_I,V_Q,V_U)$
can be organized into a $3\times 3$ covariance matrix:
\beq
{\bf M}_v\equiv\langle {\bf v}\cdot{\bf v}^\dag\rangle
=\begin{pmatrix}
C^T & C^X\cc & 0\\
C^X\cc & C^E\csq+C^B\ssq & 0\\
0 & 0 & C^E\ssq+C^B\csq
\end{pmatrix},
\label{eq:covar}
\eeq
The estimators of $C^E$ and $C^B$ are quadratic functions of 
the vector ${\bf v}$:
\beq
\hat C^E={\bf v}^\dag\cdot{\bf N}_E\cdot{\bf v},\qquad
\hat C^B={\bf v}^\dag\cdot{\bf N}_B\cdot{\bf v},
\eeq
where
\beq
{\bf N}_E=\gamma\begin{pmatrix}
0 & 0 & 0\\
0 & \csq & 0\\
0 & 0 & -\ssq
\end{pmatrix},\quad
{\bf N}_B=\gamma\begin{pmatrix}
0 & 0 & 0\\
0 & -\ssq & 0\\
0 & 0 & \csq
\end{pmatrix}.
\eeq
In this notation the proof that $\hat C^E$ is unbiased looks like
this:
\begin{subequations}
\begin{align}
\langle\hat C^E\rangle&=\langle {\bf v}^\dag \cdot {\bf N}_E\cdot
{\bf v}\rangle\\
&={\rm Tr}
\left({\bf N}_E\cdot\langle{\bf v}{\bf v}^\dag\rangle\right)
\\
&={\rm Tr}({\bf N}_E\cdot{\bf M}_v)\\
&=C^E,
\end{align}
\end{subequations}
using the explicit forms of ${\bf M}_v$ and ${\bf N}_E$ in the last step.

Throughout this section, we have made the simplifying assumption that 
the baseline ${\bf u}$ was parallel to the $x$ axis.  We now generalize
the results to the case where ${\bf u}$ points in an arbitrary direction.
Let $\alpha$ be the angle between ${\bf u}$ and the $x$ axis.  We can
calculate the optimal power spectrum estimators by rotating to a coordinate
system in which ${\bf u}$ is on the axis before applying the above 
prescription.  The rotated Stokes vector is
\beq
{\bf v}_{\rm rot}={\bf R}\cdot{\bf v}=
\begin{pmatrix}
1 & 0 & 0\\
0 & \cos 2\alpha & \sin 2\alpha \\
0 & -\sin 2\alpha & \cos 2\alpha
\end{pmatrix}\cdot{\bf v}.
\eeq
The covariance matrix (\ref{eq:covar}) applies to the rotated
vector ${\bf v}_{\rm rot}$; the original unrotated vector thus
has covariance matrix $\langle {\bf v}\cdot{\bf v}^\dag\rangle=
{\bf R}^{-1}\langle {\bf v}_{\rm rot}\cdot{\bf v}_{\rm rot}^\dag\rangle
\cdot{\bf R}=
{\bf R}^{-1}\cdot{\bf M}_v\cdot{\bf R}$.
Similarly, when applied to the unrotated data the
matrices ${\bf N}_{E,B}$ are simply replaced by ${\bf R}^{-1}\cdot{\bf N}_{E,B}
\cdot{\bf R}$.

In the next section we will continue to examine the special case of
${\bf u}$ pointing in the $x$ direction for simplicity, but these
transformations can always be made to generalize the results.

\subsection{Introduction of errors}
\label{sec:insterr}

\begin{figure*}
\includegraphics[width=3in]{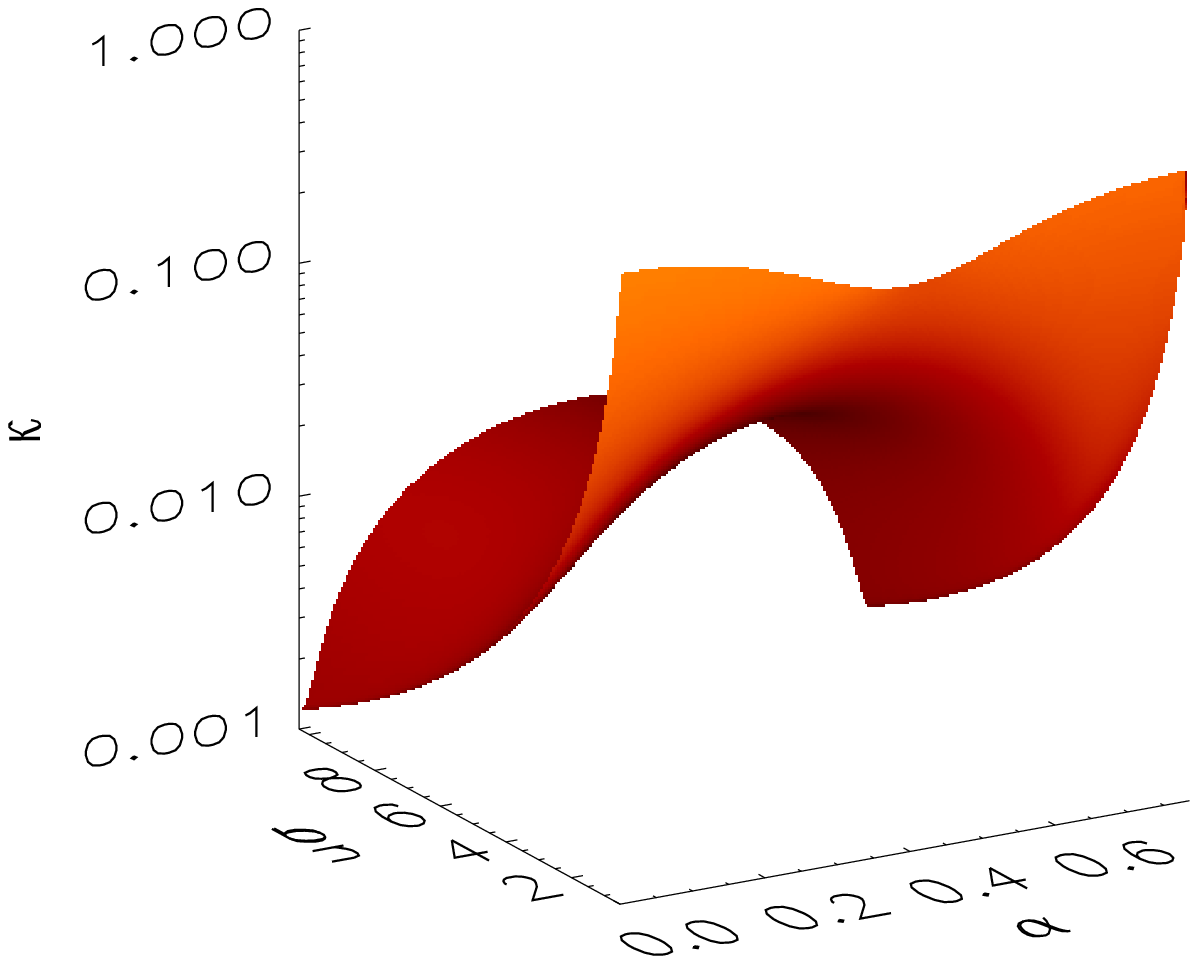}
\includegraphics[width=3in]{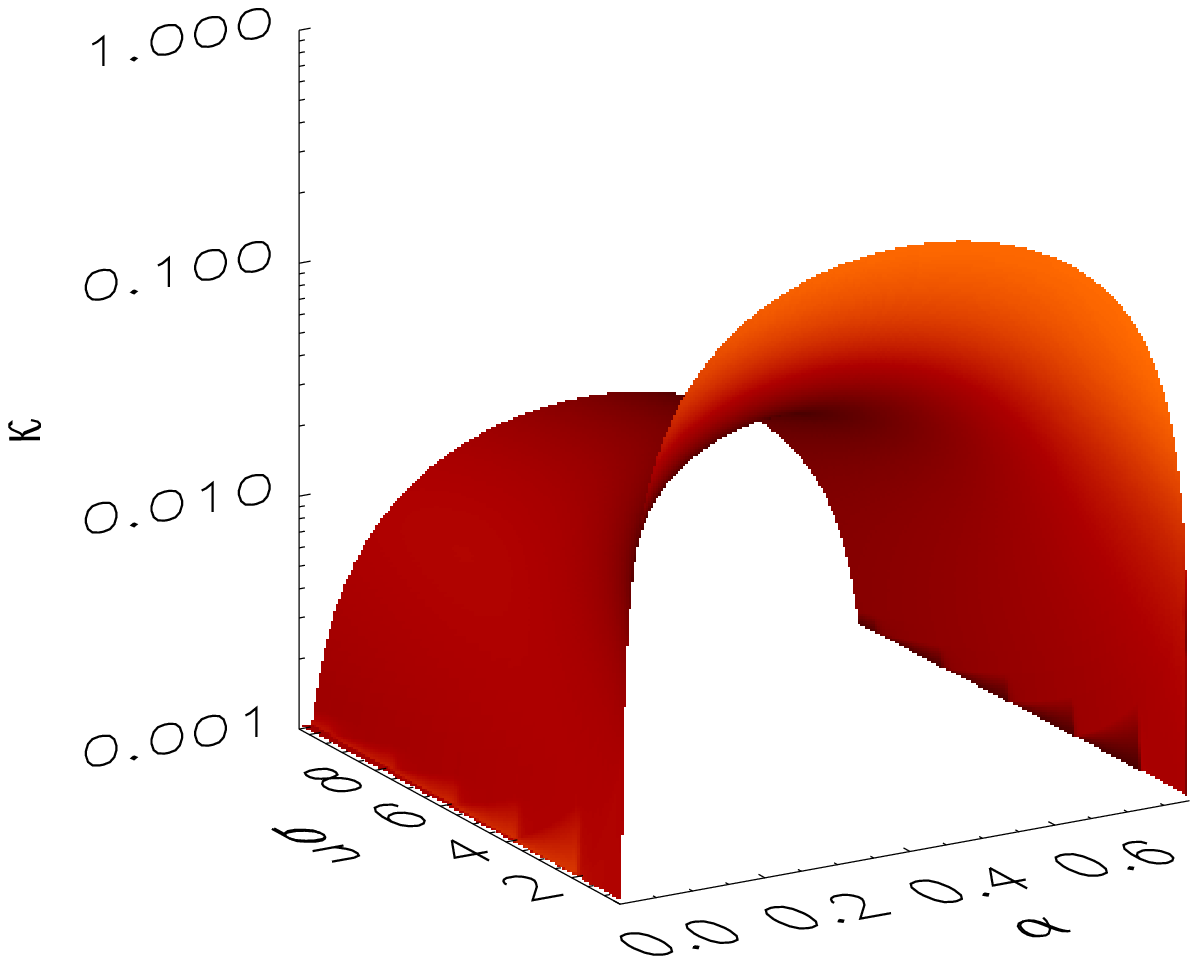}
\caption{The coefficients $\kappasub{B,EE}$ for linear-experiment
gain errors, plotted as a function of polarizer angle $\alpha$
and antenna separation $u\sigma$.  The left plot is for the parameter
$\Re(\gamma_2)$, and the right is for $\Im(\gamma_2)$.  The
coefficient for $\gamma_1$ has no $\alpha$-dependence and so is not
plotted here (but see Fig.~\ref{fig:glkappa2}).}
\label{fig:glkappa1}
\end{figure*}

Suppose that there is an error (e.g., a gain error) in the data, but
we don't know it.  Since we think we are dealing
with an error-free experiment, we 
use the optimal prescription (\ref{eq:ebest})
to estimate the power spectra from a single set
of visibilities.  
The presence of the systematic error will alter the covariances
(\ref{eq:covar}) and hence the statistical properties
of the estimators.
Let $\delta\hat C^{K}$ with $K=\{E,B\}$
be the difference between the estimate
we actually get and what we would have gotten in the absence
of systematic errors:
\beq
\delta\hat C^{K}=\hat C^{K}_{\rm actual}-\hat C^{K}_{\rm no\ errors}.
\eeq
It is natural to use the r.m.s.\ values of these differences,
\beq
\delta\hat C^{K}_{\rm rms}=\langle (\delta\hat C^{K})^2\rangle^{1/2},
\eeq
to quantify the effect of each systematic error on the power spectrum
estimates.
We focus for the moment on the case of ``instrument errors'' as described
in Section \ref{sec:inst}, deferring the generalization to beam errors
until section \ref{sec:beamresults}.

The effect of each instrument error
is to mix together the Stokes visibilities $V_I,V_Q,V_U$ in a linear
fashion (neglecting
circular polarization for the present).  If ${\bf v}$ denotes
the vector of Stokes visibilities that would be obtained
without systematic errors, the vector actually observed is
\beq
{\bf v}'={\bf v}+\delta{\bf v}={\bf v}+{\bf E}\cdot{\bf v}
\eeq
for some $3\times 3$ matrix ${\bf E}$.

The error in the power spectrum estimates is
\beq
\delta\hat C^{K}=
({\bf v}+\delta{\bf v})^\dag \cdot{\bf N}_{K}\cdot
({\bf v}+\delta{\bf v})-{\bf v}^\dag\cdot{\bf N}_{K}\cdot{\bf v}.
\label{eq:deltac}
\eeq
Assuming that the errors are small,
we can neglect the term that is quadratic in $\delta {\bf v}$:
\beq
\delta\hat C^{K}=
{\bf v}^\dag\cdot({\bf E}^\dag\cdot{\bf N}_{K}
+{\bf N}_{K}\cdot{\bf E})\cdot{\bf v}
\equiv {\bf v}^\dag\cdot{\bf A}_{K}\cdot{\bf v}.
\eeq
Assuming Gaussian fluctuations, there is a relatively simple expression
for the variance of this quantity, as shown in the Appendix:
\beq
(\delta\hat C^{K}_{\rm rms})^2=
{\rm Tr}[({\bf A}_{K}\cdot {\bf M}_v)^2]+[{\rm Tr}({\bf A}_{K}
\cdot{\bf M}_v)]^2.
\label{eq:powererror}
\eeq

As noted in the previous subsection, when these formulae are applied
to visibilities in a coordinate system that is not aligned with the
baseline vector, we must correct them by conjugating with the matrix ${\bf R}$.
All that is necessary is to replace the error matrix ${\bf E}$ with
\beq
{\bf E}\to {\bf R}\cdot{\bf E}\cdot{\bf R}^{-1}.
\eeq

For any particular source of error, we now have a recipe
for calculating the effect on the power spectrum: 
we write down
an explicit form for the matrix ${\bf E}$ and apply equation
(\ref{eq:powererror}).
The result will be a sum of terms that are quadratic
in the  band powers $C^T,C^X,C^E,C^B$.
All of these quantities
represent band powers at the same multipole $l=2\pi u$; we continue 
to omit
the multipole subscript $2\pi u$ throughout this section for simplicity.
In particular, the error in our estimate of the $B$-mode
band power generically looks like
\begin{subequations}
\begin{align}
(\delta \hat C^B_{\rm rms})^2&=\sum_{I,J}\eta^B_{IJ}C^IC^J\\
&=
\eta^B_{TT}(C^T)^2+\eta^B_{TX}C^TC^X+\eta^B_{TE} C^TC^E+\ldots  \nonumber\\
& \qquad\qquad
+\eta^B_{EE}(C^E)^2+\ldots.
\label{eq:errorexpansion}
\end{align}
\end{subequations}
Each of the coefficients $\eta$ depends on the various parameters that 
characterize the instrument errors, such as the gain fluctuations $g_i^{(j)}$
and couplings $\epsilon_i^{(j)}$.  Of course, a similar expression
would apply to $\hat C^E$.

We now imagine ``turning on'' one error at a time.
Consider a systematic error characterized by a single parameter
$p$.  The coefficients $\eta$
in the above expression will contain terms proportional to $p^2$
at leading order, because equation (\ref{eq:powererror}) is quadratic
in the error matrix ${\bf E}$:
\beq
\eta^K_{IJ}=\kappasubsq{K,IJ,p}p_{\rm rms}^2.
\label{eq:kappa}
\eeq
Here $K=\{E,B\}$ is the power spectrum we are trying to estimate; 
and $I,J=\{T,X,E,B\}$ are the power spectra being coupled to our estimate.

Equations (\ref{eq:errorexpansion}) and (\ref{eq:kappa}) together
yield the key result of this section:
\beq
(\delta\hat C_{\rm rms}^{K})^2=p_{\rm rms}^2\sum_{I,J}\kappasubsq{K,IJ,p}
C^IC^J.\tag{\ref{eq:mainresult}}
\eeq
We will drop the subscript rms on $p$ below.

For any given type of error, of course,
some of the coefficients $\eta$ will vanish.  For instance, as we saw
earlier, gain errors do not couple $I$ to $Q,U$, so there
will be no contributions coupling $C^T$ or $C^X$ to the
polarization band power estimates.  Since there is a clear hierarchy
$C^T>C^X>C^E>C^B$ (see Fig.~\ref{fig:spectra}), 
it often makes sense for each error 
to consider
only the term in (\ref{eq:mainresult})
that contains the biggest power spectra.

\begin{figure*}
\includegraphics[width=3in]{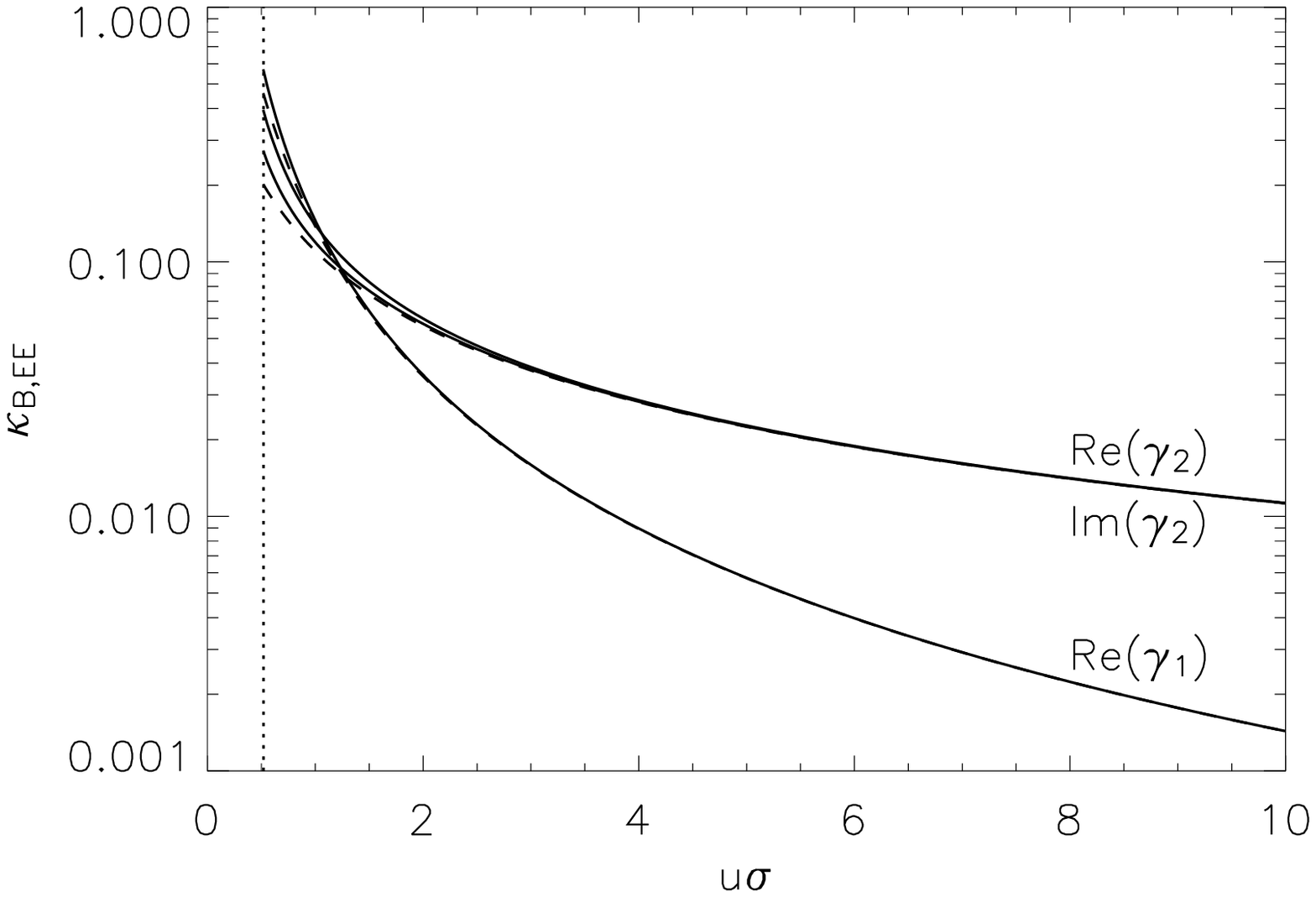}
\includegraphics[width=3in]{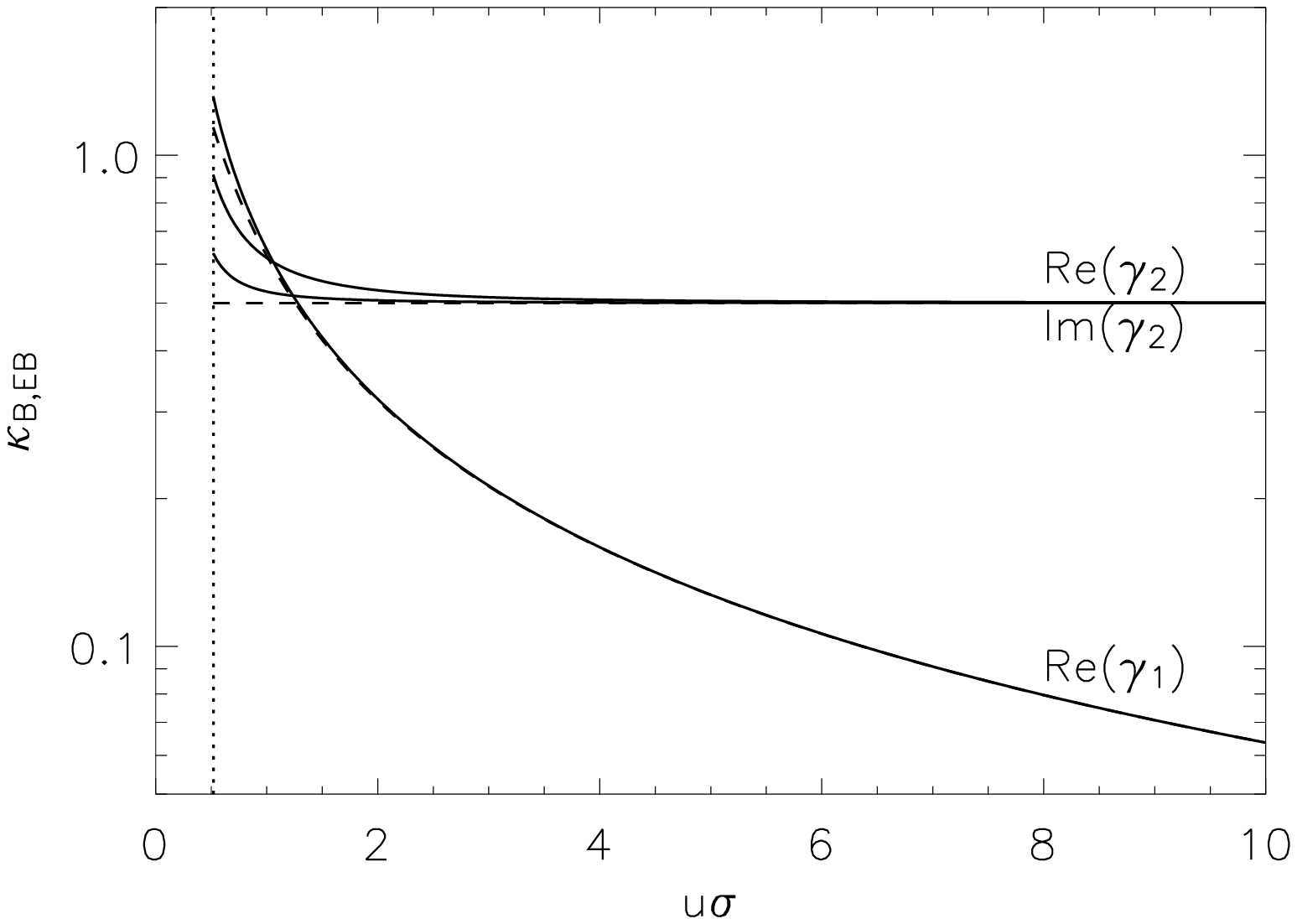}
\caption{Coefficients $\kappa$ averaged over $\alpha$ 
for linear-experiment
gain fluctuations.  The left plot shows $\kappasub{B,EE}$
for the three independent gain fluctuation parameters 
$\Re(\gamma_1),\Re(\gamma_2),\Im(\gamma_2)$ as defined in equation 
(\ref{eq:glparam}).
The right plot shows the coefficients $\kappasub{B,EB}$.
In each case, the dashed curves show the leading
term in a Taylor series in $\ssq$, which is generally a good approximation.
The vertical dotted line at $u\sigma=0.52$ corresponds to antennas
that are touching.
}
\label{fig:glkappa2}
\end{figure*}

For any one source of error,
the fractional error on the
band power will be
\beq
\frac{\delta\hat C^K_{\rm rms}}{C^K}=
p\kappasub{K,IJ,p}\frac{\sqrt{C^IC^J}}{C^K},
\label{eq:fracerror}
\eeq
assuming that one term in the sum (\ref{eq:mainresult})
dominates the error.
If we demand that this fractional error be below some specified
tolerance, then we can determine the required specification for the
input parameter $p$.
The coefficients $\kappa$ are thus
the key to assessing the severity of any particular
source of systematic error.
The next sections present calculations of these coefficients.

\section{Results: Instrument errors}
\label{sec:instresults}

This section presents the results of applying the above formalism
to the various instrument errors described in section \ref{sec:inst}.
Beam errors will be treated in the following section.
There are several cases to consider.  A table summarizing
the key results, along with a discussion of their implications,
may be found in Section \ref{sec:discuss}.

\medskip\textit{Gain errors: Linear experiment:}
Consider an experiment that measures linear polarization states,
and assume that there are gain errors $g_i^{(j)}$, ignoring couplings
($\epsilon_i^{(j)}$)
for the present.  As equation (\ref{eq:instlin}) shows, 
the resulting error in each visibility is simply proportional to the
visibility itself: $\delta V_Q=\gamma_Q V_Q$ and $\delta V_U=\gamma_U V_U$.
Here 
\beq
\gamma_{Q,U}=\tfrac{1}{2}(g_1^{(j)}+g_2^{(j)}+g_1^{(k)*}+g_2^{(k)*}),
\eeq
with the parameters $g_{i}^{(j)}$ evaluated at the time the corresponding
visibility is measured.
If $V_Q,V_U$ are measured with 
the same antennas (by rotating the polarizers in each antenna $45^\circ$), 
and if they
are measured at nearly the same time so that the gains have not drifted,
then $\gamma_Q=\gamma_U$.  It is far more likely, however, that
the two visibilities have independent gain fluctuations, in which case
$\gamma_Q$ and $\gamma_U$ should be treated as
independent,
unknown error parameters.

In the notation of the previous subsection, we can characterize these
errors with a matrix
\beq
{\bf E}=
\begin{pmatrix}
0 & 0 & 0 \\
0 & \gamma_1+\frac{1}{2}\gamma_2 & 0\\
0 & 0 & \gamma_1-\frac{1}{2}\gamma_2
\end{pmatrix},
\eeq
where 
\beq
\gamma_1=\tfrac{1}{2}(\gamma_Q+\gamma_U),
\qquad\gamma_2=\tfrac{1}{2}
(\gamma_Q-\gamma_U).
\label{eq:glparam}
\eeq

In general, we will concern ourselves only with errors
that couple larger power spectra to smaller ones.  
In an experiment to measure $E$ modes, 
gain fluctuations do not lead to any such terms.
We therefore focus on a $B$ mode experiment.
If we use this matrix to calculate the errors on
the $B$ power spectrum, the leading term is the $EE$ term:
\beq
(\delta\hat C_{\rm rms}^B)^2=\eta^B_{EE}(C^E)^2 + \ldots.
\eeq
The coefficient is
\begin{align}
\eta^B_{EE}&=
\frac{\ssq\,\csq}{2(\csq-\ssq)^2}
\bigl[|\gamma_2|^2\sin^2(4\alpha)+\nonumber\\
& 4\ssq\,\csq(4\Re(\gamma_1)^2+
(3-4\sin^2(4\alpha))\Re(\gamma_2)^2)\bigr].
\end{align}
The baseline vector ${\bf u}$ makes
an angle $\alpha$ with the $x$ axis.

The effect is characterized by three parameters
$\gamma_{1r}=\Re(\gamma_1)$, $\gamma_{2r}=\Re(\gamma_2)$, 
$\gamma_{2i}=\Im(\gamma_2)$.  
The coefficients associated with these parameters are
\begin{subequations}
\begin{align}
\kappasubsq{B,EE,\gamma_{1r}}&=\frac{8(\ssq\,\csq)^2}{(\csq-\ssq)^2},\\
\kappasubsq{B,EE,\gamma_{2r}}&=\frac{\ssq\,\csq}{2(\csq-\ssq)^2}[\sin^2(4\alpha)
+\nonumber\\
& \qquad 
4(3\cos^2(4\alpha)-\sin^2(4\alpha))\ssq\,\csq]
,\\
\kappasubsq{B,EE,\gamma_{2i}}&=\frac{\ssq\,\csq\sin^2(4\alpha)}{2(\csq-\ssq)^2}.
\end{align}
\end{subequations}

We can simplify these expressions and those to follow in two ways.
First, since a typical experiment will involve visibilities measured
with many different baseline orientations, 
we will generally average over the angle $\alpha$.  Second,
since $\ssq$ is generally a small quantity, 
we can often keep only the leading term in a Taylor expansion in $\ssq$.
In these approximations, the coefficients simplify to
\begin{subequations}
\begin{align}
\kappasub{B,EE,\gamma_{1r}}&=\sqrt{8}\ \ssq,\\
\kappasub{B,EE,\gamma_{2r}}&=\kappasub{\gamma_{2i}}=\tfrac{1}{2}\sqrt{\ssq}.
\end{align}
\end{subequations}

Figure \ref{fig:glkappa1} shows the coefficients associated with $\gamma_2$
as functions of both $\alpha$ and antenna separation $u\sigma$ (which determines
$\ssq$).  
Figure \ref{fig:glkappa2} shows the result of averaging over $\alpha$,
as well as the $\gamma_{1r}$ coefficient, which is independent of $\alpha$.  
As the figure indicates, the leading-order approximation in $\ssq$
is quite good.

We should consider whether it is always adequate to keep only the
leading term $\eta^B_{EE}$.  The full expression for the power spectrum
error contains a term $\eta^B_{EB}$ as well.  The fractional
error caused by this term scales as only $\sqrt{C^E/C^B}$ rather
than $C^E/C^B$ (eq.~\ref{eq:fracerror}).
However, the coefficient $\kappasub{B,EE}$ can
be small (see Fig.~\ref{fig:glkappa2}), especially for widely separated
antennas. 
As the right panel of Fig.~\ref{fig:glkappa2} shows, 
the coefficient $\kappa_{B,EB}$ can be much larger than $\kappa_{B,EE}$.
Determination of which term dominates must unfortunately be made
on a case-by-case basis.

\begin{figure*}
\includegraphics[width=3in]{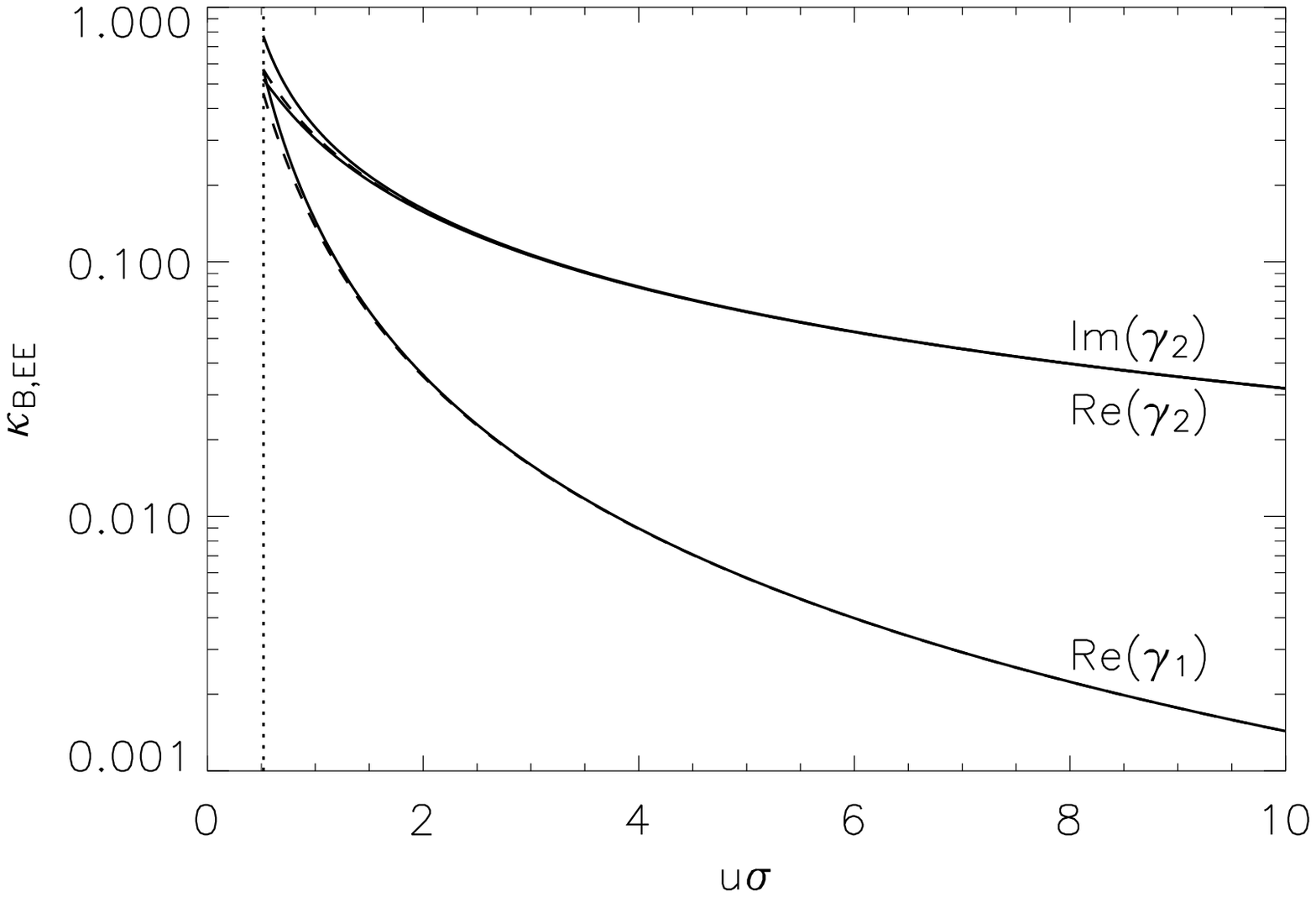}
\includegraphics[width=3in]{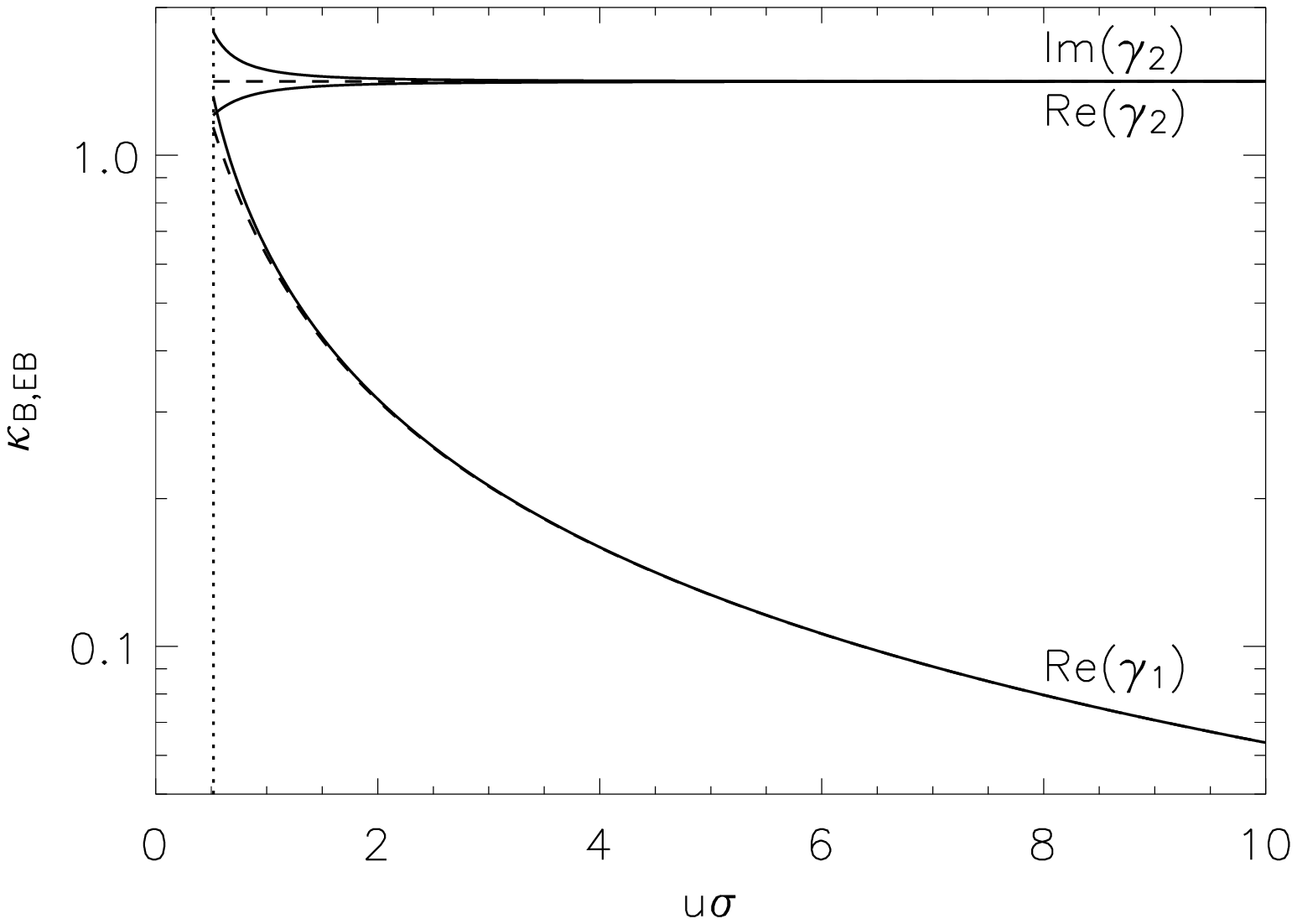}
\caption{Coefficients for gain errors in a circular-polarization experiment
with parameters defined in (\ref{eq:gcparam}).
As in Fig.~\ref{fig:glkappa2}, the left plot shows the
coefficients $\kappasub{B,EE}$, and the right shows $\kappasub{B,EB}$.
Dashed lines indicate the leading-order approximation in $\ssq$.
These coefficients are independent of $\alpha$, so no averaging was necessary.
The vertical dotted line corresponds to antennas that are touching.
}
\label{fig:gc}
\end{figure*}

The size of the coefficient $\kappa$ is largely determined by its dependence
on the small parameter $\ssq$.  As a general rule, we need to weigh
the importance of a subdominant contribution to the error (i.e., one
that ranks lower in the $T,X,E,B$ hierarchy) if that error has a weaker
dependence on $\ssq$.  In the case of the parameter $\gamma_1$,
for example, the EE term has $\kappasub{B,EE}\propto \ssq$ but
$\kappasub{B,EB}\propto\sqrt{\ssq}$.  For $\gamma_2$, the EB coefficient
has an $\ssq$-independent term.
Thus for small $\ssq$ (large separation),
the EB terms may become more important than the EE terms.  
See Section \ref{sec:discuss}
and especially Table \ref{table}
for further discussion of this point.

\medskip\textit{Gain errors: Circular experiment:}
In this case, we assume an experiment that measures $V_Q$ and $V_U$
simultaneously by interfering right and left circular polarizations.
Again, we focus on an experiment to measure $B$ modes, as gain errors
do not pose a serious problem in an $E$ mode experiment.
From equations (\ref{eq:instcirc}), 
we see that gain errors $g_i^{(j)}$ produce an error matrix of the form
\beq
{\bf E}=\begin{pmatrix}
0 & 0 & 0\\
0 & \gamma_1 & i\gamma_2 \\
0 & -i\gamma_2 & \gamma_1
\end{pmatrix},
\eeq
where 
\begin{subequations}
\label{eq:gcparam}
\begin{align}
\gamma_1&=\tfrac{1}{2}(g_1^{(j)}+g_2^{(j)}+g_1^{(k)*}+g_2^{(k)*}),\\
\gamma_2&=\tfrac{1}{2}(g_1^{(j)}-g_2^{(j)}-g_1^{(k)*}+g_2^{(k)*}).
\end{align}
\end{subequations}
As in the previous case, the dominant error contribution to a measurement
of $B$ power is the $EE$ term:
\beq
\eta_{EE}^B=\frac{2\ssq\,\csq(|\gamma_2|^2+4\ssq\,\csq(\Re(\gamma_1)^2-
\Re(\gamma_2)^2))}{(\csq-\ssq)^2}.
\eeq
The coefficients are
\begin{subequations}
\begin{align}
\kappasubsq{B,EE,\gamma_{1r}}&=
\frac{8(\ssq\,\csq)^2}{(\csq-\ssq)^2}\approx 8(\ssq)^2,\\
\kappasubsq{B,EE,\gamma_{2r}}&=\frac{2\ssq\,\csq(1-4\ssq\,\csq)}{(\csq-\ssq)^2}
\approx 2\ssq,\\
\kappasubsq{B,EE,\gamma_{2i}}&=\frac{2\ssq\,\csq}{(\csq-\ssq)^2}\approx 2\ssq,
\end{align}
\end{subequations}
where the approximate equalities are the leading terms in an expansion
in $\ssq$.  
See Figure \ref{fig:gc}.

As in the case of a linear polarization experiment, the $EB$
error term can become dominant for widely-separated antennas.  
In particular, for
the parameter $\gamma_2$ the EB coefficient has a term independent
of $\ssq$:
$\kappasub{B,EB,\gamma_{2r,i}}=\sqrt{2}$ to leading order in $\ssq$.

\medskip\textit{Couplings:}
Next we turn to errors parameterized by the ``coupling'' terms
$\epsilon_i^{(j)}$ in the instrument Jones matrix.  These errors include
electronic cross-talk as well as errors in the alignments of the polarizers
in a linear experiment.
In both linear and circular experiments, these errors couple $I$ into
$Q,U$, so the dominant terms will be those involving the temperature
power spectrum.  In this section, unlike the previous ones, we should
consider $E$ as well as $B$ mode experiments.

The error matrix characterizing $I\to Q,U$ leakage in this situation
is
\beq
{\bf E}=\begin{pmatrix}
0 & 0 & 0\\
\varepsilon_1 & 0 & 0 \\
\varepsilon_2 & 0 & 0
\end{pmatrix}.
\eeq
Here $\varepsilon_1,\varepsilon_2$ are 
the coefficients of $\Vo_I$ in equations
(\ref{eq:instlin}) and (\ref{eq:instcirc}):
For a linear experiment,
\beq
\varepsilon_{1,2}=\tfrac{1}{2}(\epsilon_1^{(j)}+\epsilon_2^{(j)}+
\epsilon_1^{(k)*}+\epsilon_2^{(k)*}),
\label{eq:coupleparam1}
\eeq
with the parameters $\epsilon_i^{(j)}$ evaluated when the corresponding
visibility is measured.  For a circular experiment, 
\begin{subequations}
\label{eq:coupleparam2}
\begin{align}
\varepsilon_{1}&=\tfrac{1}{2}(\epsilon_1^{(j)}+\epsilon_2^{(j)}+
\epsilon_1^{(k)*}+\epsilon_2^{(k)*}),\\
\varepsilon_{2}&=\tfrac{1}{2}(-\epsilon_1^{(j)}+\epsilon_2^{(j)}+
\epsilon_1^{(k)*}-\epsilon_2^{(k)*}).
\end{align}
\end{subequations}
In the linear case, there are also terms that couple $Q$ and $U$.
We omit these, as the errors they produce are always small
in comparison to the terms involving $I$.

\begin{figure}
\includegraphics[width=3in]{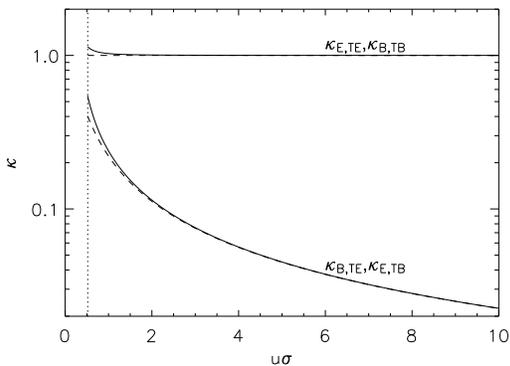}
\caption{Coefficients for coupling errors parameterized as in equations
(\ref{eq:coupleparam1}) and (\ref{eq:coupleparam2}). 
Dashed and dotted curves are as in the previous figures.
}
\label{fig:cross}
\end{figure}

Consider first a $B$ mode experiment.
The $TT$ and $TX$ terms in (\ref{eq:errorexpansion}) vanish, so
the dominant contribution is the $TE$ term.  After averaging over $\alpha$,
this term is
\beq
\eta_{TE}^B=\frac{\ssq\,\csq(|\varepsilon_1|^2+|\varepsilon_2|^2)}{
(\csq-\ssq)^2},
\eeq
so the coefficients for the parameters $\Re(\varepsilon_1),\Im(\varepsilon_1),
\Re(\varepsilon_2),\Re(\varepsilon_2)$ are
\beq
\kappasubsq{B,TE,\varepsilon}=\frac{\ssq\,\csq}{(\csq-\ssq)^2}\approx\ssq,
\eeq
as shown in Figure \ref{fig:cross}.
The $TB$ term has an $\ssq$-independent piece:
\beq
\kappasubsq{B,TB,\varepsilon}=\frac{1+3\ssq\,\csq}{(\csq-\ssq)^2}
\approx 1,
\eeq
which can be important for large antenna
separation.  See Table \ref{table}.

The coefficients for an $E$ mode experiment are the
same as for a $B$ mode experiment with $E$ and $B$ switched.
The dominant contribution is therefore $\kappasub{E,TE}\approx 1$.

\section{Beam errors}
\label{sec:beamresults}

In section \ref{sec:insterr}, we derived a method of forecasting
the effects of instrument errors on power spectrum estimates.  We
now generalize this method to the case of beam errors.

In the case of instrument errors, we were able to write the
errors in the visibilities as $\delta{\bf v}={\bf E}\cdot{\bf v}$,
where the error matrix ${\bf E}$ depended only on the unknown error
parameters.  Beam errors cannot be treated in
this way.
Both
$\delta{\bf v}$ and ${\bf v}$ are integrals over the Stokes
parameters, but with different weightings in Fourier space.  As a result,
one
is not a simple linear transformation of the other. 
Fortunately, for a number of important sources of error, the differences
in Fourier-space weighting are modest: the errors sample
roughly if not exactly the same regions of the Fourier plane.
We can therefore still express our final results in the form
(\ref{eq:mainresult}), after we have made some adaptation to
the formalism of section \ref{sec:insterr}.

Combine ${\bf v}$ and $\delta{\bf v}$ together into a 6-dimensional
vector ${\bf w}=({\bf v},\delta{\bf v})=(V_I,V_Q,V_U,\delta V_I,\delta V_Q,
\delta V_U)$.  To leading order in $\delta{\bf v}$,
the error in the power spectrum estimate is
\beq
\delta\hat C^{K}={\bf w}^\dag\cdot {\cal N}_{K}\cdot{\bf w},
\eeq
where $K=\{E,B\}$ and the matrix ${\cal N}$ can be written in block form as
\beq
{\cal N}_{K}=\begin{pmatrix}0 & {\bf N}_{K}\\
{\bf N}_{K} & 0\end{pmatrix},
\eeq
with ${\bf N}_{K}$ the same as for instrument errors.
It is straightforward to check that this reduces to (\ref{eq:deltac}).

Using the identity proved in the Appendix again, we can write
\beq
(\delta\hat C^{K}_{\rm rms})^2=\tr\left[({\cal N}_{K}\cdot {\bf M}_w)^2
\right]+\left[\tr({\cal N}_{K}\cdot{\bf M}_w)\right]^2,
\eeq
where ${\bf M}_w=\langle{\bf w}\cdot{\bf w}^\dag\rangle$ is the 
covariance matrix of the vector ${\bf w}$.  We now need a recipe
for calculating the elements of this covariance matrix, which 
will contain terms proportional to the various power spectra.

In sections \ref{sec:vis}
and \ref{sec:beam}, we expressed
each component of ${\bf v}$ and $\delta{\bf v}$ as an integral
over the Stokes parameters.  To be explicit, let ${\bf s}({\bf k})
=(\tilde I({\bf k}),\tilde Q({\bf k}),\tilde U({\bf k}))$ be a vector
giving the Fourier-space Stokes parameters.  Each component of ${\bf w}$
can be expressed in the form 
\beq
w_i=\int d^2k\,{\bf W}_i({\bf k})\cdot{\bf s}({\bf k})
\eeq
for some vector-valued window function ${\bf W}_i$.  A covariance
matrix $\langle w_iw_j^*\rangle$ then becomes an integral over ${\bf k}$
of the two window functions times the covariances
of the Stokes parameters $\langle {\bf s}\cdot{\bf s}^\dag\rangle$.
The latter are proportional to the input power spectra.  So once
we have written down the window functions for the visibilities
and their associated systematic errors, we can calculate
an expression giving contributions
to the error $\delta\hat C^K_{\rm rms}$ in terms of the input power spectra,
just as in the case of instrument errors.
For some parameterizations of beam errors, the resulting integrals
can be performed analytically to yield closed-form expressions
like those in the previous section for the coefficients $\kappa$.
However, the resulting expressions are complicated and unenlightening,
so we present the results of numerical integration instead.

As in the previous section, we now examine detailed case-by-case results.
Section \ref{sec:discuss} provides a summary of the implications.

As noted earlier, 
the set of possible forms for beam errors is dauntingly large.
Our treatment will necessarily be
restricted to a small set of physically motivated possibilities
rather than exploring the entire space.  As in
the previous section, we will imagine turning on one error at a time.

\medskip
\textit{Differential pointing errors (``squint''):}
Suppose that some antennas have slight pointing offsets relative to 
others.  This situation can be treated as a beam mismatch error
as in equation (\ref{eq:mismatch}).

Let $A_0(\rhat)$ be the antenna pattern in the absence of the pointing
errors, which we will take to be a Gaussian.  Then in the notation
of 
equation (\ref{eq:mismatch}),
the antenna pattern for the $j$th antenna is 
\beq
A^{(j)}(\rhat)=A(\rhat+\delta\rhat_j),
\eeq
where $\delta\rhat_j$ is the pointing error of the $j$th antenna.
Using equations (\ref{eq:beamvis}), we find that each visibility
looks like
\beq
V_Z=\int d^2\rhat e^{-2\pi i{\bf u}_{jk}\cdot\rhat}Z(\rhat)A^{(j)}(\rhat)
A^{(k)*}(\rhat),
\eeq
where $Z=\{Q,U\}$.
The product of two Gaussians is a Gaussian centered at the
midpoint of the two:
$A^{(j)}(\rhat)
A^{(k)}(\rhat)\propto\exp
[-(\rhat-\frac{1}{2}(\delta\rhat_j+\delta\rhat_k)^2)/(2\sigma^2)]$.
That is, each visibility is calculated using an effective beam pattern
that is shifted by the average of the shifts of the two antennas.
For any given antenna pair $(jk)$, we define an error parameter
\beq
\bm{\delta}_{jk}=\frac{\delta\rhat_j+\delta\rhat_k}{2\sigma},
\eeq
the average of the two antennas' pointing errors in units of the beam width.

\begin{figure}
\includegraphics[width=3in]{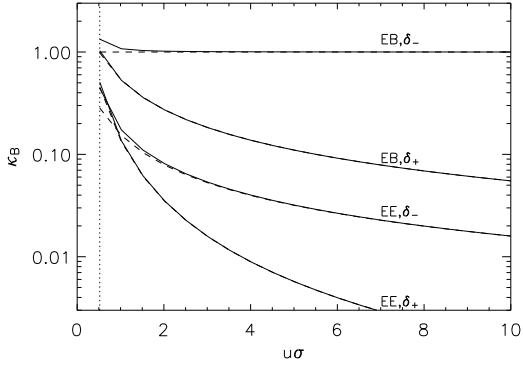}
\caption{Error coefficients for beam pointing errors.  The quantity
$\bm{\delta}_+$ is a common pointing error (the same for both $V_Q,V_U$),
while $\bm{\delta}_-$ is a relative pointing error.  Both
$\bm{\delta}_\pm$ are measured in units of the beam width.  Results
are averaged over directions of $\bm{\delta}_\pm$.  Furthermore,
in the case of $\bm{\delta}_-$, an average has
been taken over the angle $\alpha$ between ${\bf u}$ and the $x$ axis.
Dashed and dotted curves are as in the previous figures.}
\label{fig:squint}
\end{figure}

Shifting a function by an amount $\bm{\Delta}$ is equivalent to
multiplying its Fourier transform by $e^{i{\bf k}\cdot\bm{\Delta}}$, so 
by (\ref{eq:visfourier}) the visibility is
\beq
V_Z=\int d^2k\,\tilde Z({\bf k})\widetilde{A_0^2}({\bf k}-2\pi{\bf u})^*
e^{-i({\bf k}-2\pi{\bf u})\cdot\bm\delta_{jk}\sigma}.
\eeq
To leading order in $\bm{\delta}_{jk}$, the error is
\beq
\delta V_Z=
-i\int d^2k\,\tilde Z({\bf k})\widetilde{A_0^2}({\bf k}-2\pi{\bf u})^*
({\bf k}-2\pi{\bf u})\cdot\bm{\delta}_{jk}\sigma.
\eeq

As before, we imagine a single measurement pair $(V_Q,V_U)$ corresponding
to the same baseline ${\bf u}$.  Let $\bm{\delta}_Q$ be the value
of $\bm{\delta}_{jk}$ corresponding to the visibility $V_Q$ and
$\bm{\delta}_U$ be the value corresponding to $V_U$.  In a circular
experiment, the two visibilities are measured with the same antenna pair,
so $\bm{\delta}_Q=\bm{\delta}_U$, while in a linear experiment
they should be regarded as independent error parameters.  It is convenient
to express our final results in terms of the sum and difference,
\beq
\bm{\delta}_{\pm}=\tfrac{1}{2}(\bm{\delta}_Q\pm\bm{\delta}_U).
\label{eq:squintparams}
\eeq
In a circular experiment, $\bm{\delta}_-=0$.

We can now calculate the various correlations $\langle \delta V_Q V_Q\rangle$,
etc.,
by integrating over ${\bf k}$.  The result
will contain terms proportional to the band powers $C_{2\pi u}^E,C_{2\pi u}^B$,
and quadratic in the parameters $\bm{\delta}_+,\bm{\delta}_-$.
We can therefore define parameters $\kappa$ characterizing
these errors exactly as in the case of instrument errors 
[equation (\ref{eq:mainresult})].

As in the previous cases, the severity of the errors depends on $\ssq$, which
characterizes the degree of $EB$ mixing within
each visibility pair.  Furthermore, the components of $\bm{\delta}_\pm$
parallel and perpendicular to the baseline, 
$\delta_{\pm\parallel }$
and $\delta_{\pm\perp}$, contribute differently.  
Finally, in the case of $\bm{\delta}_-$, the results depend on the
angle $\alpha$ between the baseline and the coordinate axis.
For simplicity, we have averaged over both components $\delta_\parallel$
and $\delta_\perp$ for each of $\bm{\delta}_\pm$, and we have also 
averaged over $\alpha$.
Figure \ref{fig:squint} illustrates the
resulting coefficients.
As before, the coefficients are well approximated by the leading-order
terms in an expansion in $\ssq$, which are given in Table \ref{table}.
Not surprisingly, 
$\bm{\delta}_-$ is a greater source of error than $\bm{\delta}_+$.
As comparison with Figures \ref{fig:glkappa2} and \ref{fig:gc} indicates,
the effects of differential pointing errors ($\bm{\delta}_-$)
are generally similar to those of gain errors.

\medskip\textit{Beam shape errors:}
Equation (\ref{eq:mismatch}) can also be used to model errors
in the beam shape.  To illustrate this, we consider Gaussian
beams with errors in the beam width.

Assume that in an ideal, error-free experiment all antennas have
azimuthally symmetric Gaussian beam patterns with beam width $\sigma$.
Suppose that in actuality each antenna has an elliptical
beam pattern with different beam widths $\sigma_1^{(j)},\sigma_2^{(j)}$
along its two principal axes.
As equation (\ref{eq:beamvis}) indicates, the effective beam pattern
for each visibility $V^{(jk)}$ is just the product of the two antenna
patterns.  The product of Gaussians is a Gaussian, so the effective
visibility beam pattern will be of the form
\beq
A^{(j)}(\rhat)A^{(k)}(\rhat)\propto
\exp[{-\rhat\cdot(\bm{1}+\bm{\Delta}_{jk})\cdot\rhat/(2\sigma^2)}]
.
\label{eq:beamwidth}
\eeq
Here the symmetric $2\times 2$ matrix $\bm{\Delta}_{jk}$ characterizes
the deviation of the beam from the ideal symmetric Gaussian of width $\sigma$.
To be specific, the eigenvectors of $\bm{\Delta}_{jk}$ give the two
principal axes of the elliptical beam, and the beam widths are 
$\sigma/\sqrt{1+\lambda_i}$ where $\lambda_1,\lambda_2$ are the eigenvalues.
We will assume that the errors are small and work to leading
order in $\lambda_i$.  The
fractional errors in the beam width in the two principal directions are then
$\delta\sigma_i/\sigma=-\lambda_i/2$,
where $i=1,2$ label the two principal axes of the beam.

As usual we consider a visibility pair $(V_Q,V_U)$ measured with a common
baseline ${\bf u}$.  In the case of a linear experiment, the two visibilities
may be measured with different antenna pairs, so we should consider
two sets of beam shape parameters characterized by matrices $\bm{\Delta}_Q,
\bm{\Delta}_U$.  In a circular experiment where both visibilities
are measured simultaneously with a single antenna pair, $\bm{\Delta}_Q=
\bm{\Delta}_U$.  As we have seen before, we can
treat both cases simultaneously by defining
\beq
\bm{\Delta}_\pm=\tfrac{1}{2}(\bm{\Delta}_Q\pm\bm{\Delta}_U).
\eeq
The matrix $\bm{\Delta}_+$ characterizes the average beam shape when
the two visibilities are measured, and $\bm{\Delta}_-$ characterizes
errors in beam shape that differ between $V_Q,V_U$.  In both cases, the
two eigenvalues of the matrices give fractional errors in beam width
in the two principal directions: 
\beq
\zeta_{\pm,i}\equiv
\frac{\delta\sigma_{\pm,i}}{\sigma}=-\frac{\lambda_{\pm,i}}{2}.
\label{eq:beamshapeparams}
\eeq
Here $i=1,2$ labels the two principal axes for each of $\bm{\Delta}_\pm$.

\begin{figure}
\includegraphics[width=3in]{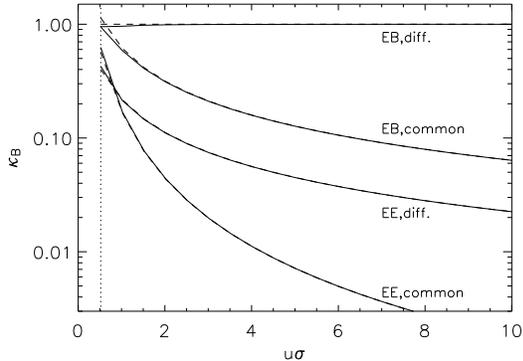}
\caption{Error coefficients for beam shape errors, parameterized
by the coefficients $\zeta_{+,i}$ (``common'') and $\zeta_{-,i}$ (``diff.'')
in equation (\ref{eq:beamshapeparams}).
Dashed and dotted curves are as in previous figures.
}\label{fig:beamshape}
\end{figure}

We will refer to errors parameterized by $\zeta_{+,i}$ as {\it common}
beam shape errors and to those parameterized by $\zeta_{-,i}$ as
{\it differential} errors.
For a circular experiment we expect differential errors to vanish,
while for a linear experiment both should be of comparable magnitude.
We will consider separately the effects of common and differential
errors.

In each of the two cases, 
there are three parameters: $\zeta_1,\zeta_2$, and an angle $\beta$
giving the orientation of the principal axes relative to the
coordinate axes used to define $Q,U$.  For common errors, the results
are independent of $\beta$, but for differential errors they depend
on $\beta$ (unless $\zeta_{-,1}=\zeta_{-,2}$, in which case there is
rotational symmetry).  We assume that the principal axes are randomly
oriented, so we average over $\beta$ in the results below.

For both common and differential errors,
there are two qualitatively different possibilities
one might wish to consider.  If $\zeta_1=\zeta_2$, then the beam
is circular, and we have made an error only in its width.  On the
other hand, the case $\zeta_1=-\zeta_2$ corresponds to a pure 
beam shape error, with the beam stretched along one axis and squeezed
equally along the other.  Of course, the most general case would
be a combination of the two.  The final results (after averaging
over $\beta$ where appropriate) turn out to be the
same in both cases: the error depends only on the combination
$\zeta_1^2+\zeta_2^2$ regardless of the relative signs.

Figure \ref{fig:beamshape} shows the coefficients $\kappa$ associated with
beam shape errors.  The results are quite similar to those for pointing
errors.  In particular, the differential errors that arise in a linear
experiment are more severe than the common errors, which
arise in both linear and circular experiments.

\medskip

\begin{figure}
\includegraphics[width=3in]{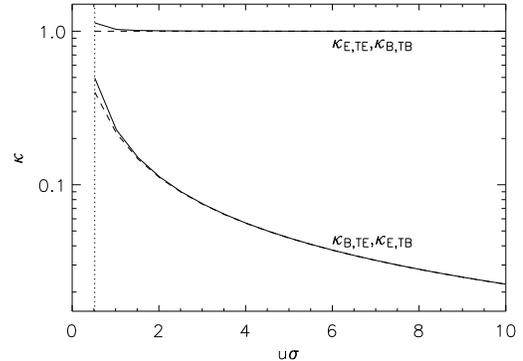}
\caption{Coefficients for cross-polar beam response, parameterized
by $\mu_{Q},\mu_U$ [eq. (\ref{eq:xpparams})].
Dashed and dotted curves are as in previous figures.}
\label{fig:xp}
\end{figure}

\begin{table*}
\begin{ruledtabular}
\begin{tabular}{ccccccc}
{\bf Experiment} & {\bf Measurement} & {\bf Error} & {\bf Dominant}
& {\bf Fiducial} & {\bf Secondary} & {\bf Fiducial}\\
{\bf type}& & {\bf source} & {\bf contribution} & $\delta\hat C/C$ &
{\bf contribution} & $\delta\hat C/C$\\
\noalign{\hrule}
\\[-8pt]
Linear & 
$B$ & 
Gain error\footnote[1]{Eq. (\ref{eq:glparam}).} & 
$\kappasub{B,EE,\gamma_{2}}=\frac{1}{2}\sqrt{\ssq}$ & 
$21\gamma_{2}$ & 
$\kappasub{B,EB,\gamma_{2}}=\frac{1}{2}$ & $8.7\gamma_{2}$
\\
Circular &
 $B$ &
 Gain error\footnote[2]{Eq. (\ref{eq:gcparam}).} &
 $\kappasub{B,EE,\gamma_2}=\sqrt{2\ssq}$ &
$60\gamma_2$ &
 $\kappasub{B,EB,\gamma_2}=\sqrt{2}$ & $24\gamma_2$
\\
Linear/Circular &
 $B$ &
 Coupling\footnote[3]{Eq. (\ref{eq:coupleparam1}) (linear); Eq. 
(\ref{eq:coupleparam2}) (circular).}&
 $\kappasub{B,TE,\varepsilon}=\sqrt{\ssq}$ &
$730\varepsilon$ &
 $\kappasub{B,TB,\varepsilon}=1$ & $300\varepsilon$
\\
Linear/Circular &
 $E$ &
 Coupling\footnotemark[3] &
 $\kappasub{E,TE,\varepsilon}=1$ &
$17\varepsilon$ &
 --- &
 ---
\\[6pt]
\noalign{\hrule}
\\[-8pt]
Linear & 
$B$ & 
Pointing\footnote[4]{Eq. (\ref{eq:squintparams})} &
$\kappasub{B,EE,\delta_{-}}=\sqrt{\ssq/2}$ &
$30\delta_-$ & 
$\kappasub{B,EB,\delta_{-}}=1$ & 
$17\delta_{-}$
\\
Circular & 
$B$ & 
Pointing\footnotemark[4] &
$\kappasub{B,EE,\delta_{+}}=\sqrt{8}\ssq$ &
$17\delta_+$ & 
$\kappasub{B,EB,\delta_{+}}=\sqrt{6\ssq}$ & 
$6\delta_{+}$
\\
Linear & 
$B$ & 
Beam shape\footnote[5]{Eq. (\ref{eq:beamshapeparams})} &
$\kappasub{B,EE,\zeta_{-}}=\sqrt{\ssq}$ &
$42\zeta_-$ & 
$\kappasub{B,EB,\zeta_{-}}=1$ & 
$17\zeta_{-}$
\\
Circular & 
$B$ & 
Beam shape\footnotemark[5] &
$\kappasub{B,EE,\zeta_{+}}=3.5\ssq$ &
$21\zeta_+$ & 
$\kappasub{B,EB,\zeta_{+}}=\sqrt{8\ssq}$ & 
$7\zeta_{+}$
\\
Linear/Circular&
$B$ & 
Cross-polarization\footnote[6]{Eq. (\ref{eq:xpparams})} & 
$\kappasub{B,TE,\mu}=\sqrt{\ssq}$ &
$730\mu$ & 
$\kappasub{B,TB,\mu}=1$ & 
$300\mu$
\\
Linear/Circular&
$E$ & 
Cross-polarization\footnotemark[6] & 
$\kappasub{E,TE,\mu}=1$ &
$17\mu$ & 
--- & 
---
\\[6pt]
\end{tabular}
\end{ruledtabular}
\caption{Effects of instrument errors (above line) and beam
errors (below line). See Section \ref{sec:discuss}
for details.}
\label{table}
\end{table*}

\textit{Cross-polarization:}
The final case we consider is azimuthally-symmetric cross-polarization,
with antenna patterns of the form (\ref{eq:xpantenna}).  We consider
an ideal
experiment to be one with cross-polar terms $A_1^{(i)}=0$ for all antennas.
The error term can in principle be an arbitrary function of $r$.  We
generally expect cross-polar response to be small near the beam center,
so we adopt the following simple form for the cross-polar response:
\beq
A_1^{(i)}(r)=\mu_i\frac{r^2}{\sigma^2}A_0(r),
\label{eq:xpform}
\eeq
where $A_0$ is assumed to have the usual Gaussian form and $\mu_i$
is the parameter characterizing the size of the error.  

As an aside, note that
this particular form arises in one simple model of an antenna.
Suppose the antenna lies in the $xy$ plane and responds equally to 
both $x$ and $y$ components of the incoming electric field,
with no sensitivity to the $z$ component.  In the flat-sky limit
such an antenna has no cross-polar response, but sky curvature
introduces cross-polarization of this form (because $E_\theta$
is reduced a factor of $\cos\theta\approx 1-\frac{1}{2}\theta^2$ upon
projection onto the $xy$ plane, while $E_\phi$ is unchanged).
This cross-polarization is characterized by $\mu=\frac{1}{2}\sigma^2$
with $\sigma$ in radians.  (Incidentally, when sky curvature
is taken into account one must be careful to distinguish
among inequivalent definitions
of ``cross-polarization.''  The most natural one in this context,
because it respects azimuthal symmetry, is ``definition 3'' in 
ref.~\cite{ludwig}.)

The relevant quantity for characterizing the error in each
of the visibilities $V_Q,V_U$ is
\beq
\mu_{Q,U}=\tfrac{1}{2}(\mu_j+\mu_k),
\label{eq:xpparams}
\eeq
the average of the two $\mu$ parameters when each of $V_Q,V_U$ is measured.
As usual, for a circular experiment $\mu_Q=\mu_U$ while in a linear
experiment the two are independent.  In this case, however, 
it makes no difference which case we consider, as the error contributions
due to $\mu_Q,\mu_U$ simply add independently (in quadrature).

Figure \ref{fig:xp} shows the leading error coefficients for 
this case.  Since these
errors couple $I$ into $Q,U$, the dominant terms are those
involving the temperature power spectrum, and errors can be 
quite significant for both $E$ and $B$ measurements.

\section{Discussion}
\label{sec:discuss}

This paper has presented a method of quantifying the effects of a variety
of systematic errors on estimates of the CMB polarization power spectra
and have applied the method to a variety of possible errors.
Let us begin by summarizing these results in a more compact form.

To illustrate the relative magnitudes of the various sources of error,
let us consider a fiducial set of experimental parameters.  Let us assume
that the true power spectra in the range of multipoles probed
by our experiment are in the ratio
\beq
C^T:C^E:C^B = 300^2:300:1,
\eeq
roughly typical for subdegree-scale experiments.  Furthermore, let us assume
a fiducial value of 
\beq
\ssq=0.02,
\eeq
which corresponds roughly to a baseline formed by a pair of antennas
separated by three times the antenna diameter.

Having chosen these fiducial values we can work out the effect of
any particular error source.  For instance, consider the effect
of gain errors on a linear experiment
aiming to measure $B$ polarization.  The leading contribution
to the error is the one that couples $EE$ to $B$, with
\beq
\kappasub{B,EE,\gamma_2}=\tfrac{1}{2}(\ssq)^{1/2}=0.071.
\eeq
The effect on the measurement of $C^B$ is
\beq
\frac{\delta\hat C^B_{\rm rms}}{C^B}=\kappasub{B,EE,\gamma_2}\gamma_2
\frac{C^E}{C^B}=21\gamma_2.
\eeq
Say for instance that we wish systematic errors to have at most a 10\%
effect.  Then $21\gamma_2<0.1$ or $\gamma_2<5\times 10^{-3}$.
Of course $\gamma_2$ here represents the r.m.s.\ value of an unknown
gain fluctuation, so this should be interpreted as an estimate of the level
to which gain fluctuations must be understood.

Table \ref{table} summarizes the results of such calculations
for the various errors considered in this paper.  A horizontal
line separates instrument from sky errors.  In each case, the
dominant term listed is the one that involves the largest input power
spectra.  In cases where $\ssq$ is small, an error
term that is lower in the hierarchy may be of comparable significance to
the dominant term.  The table therefore lists a second contribution
to each error where appropriate.  
This second contribution has $\kappa$ more weakly
dependent on $\ssq$ than the dominant contribution, so for large antenna
separation it may be the more important term (although
for the fiducial parameters adopted here it never is).  
In the cases of coupling
errors and cross-polarization in an $E$-mode measurement, the
dominant term is independent of $\ssq$, so there is no need to consider
a second term.

In all entries in the table, the coefficients are averaged over $\alpha$
and calculated with the leading-order term in an expansion in $\ssq$,
As figures \ref{fig:glkappa2}-\ref{fig:xp}
indicate, the latter approximation
is excellent.

In all cases, the error parameters should be taken as r.m.s.\ residuals
after known errors have been removed.  For instance, as 
we noted in the previous section, sky curvature can induce
cross-polarization characterized by $\mu=\frac{1}{2}\sigma^2$.
Presumably that effect would be known and accounted for;
the parameter $\mu$ 
in Table \ref{table} represents an unknown and hence unmodeled
additional component.

Not surprisingly, the coupling parameters $\varepsilon$ and cross-polarization
$\mu$ are of the greatest
concern, since they couple the temperature power spectrum to polarization
measurements.  in particular, if we want $\delta\hat{C}^B/C^B$ to be, say
at most 10\%, then these parameters must be $\varepsilon,\mu\lesssim
10^{-4}$.  

Recall that for a linear experiment the coupling parameters
can be used to describe errors
in the alignment of the polarizers, so a $B$ mode experiment would require
alignment with a precision $\sim 10^{-4}$ radians or $\sim 0.3'$.
For the $E$ power spectrum, on the other hand, the required tolerance
is about $0.3^\circ$.

For pointing and beam shape errors, circular experiments have
an advantage over linear experiments, because errors
that differ between measurement of $V_Q$ and $V_U$ (parameterized
by $\delta_-,\zeta_-$) are absent.  Gain errors, on the other hand, are
worse in a circular experiment.

All of the errors in Table \ref{table}
are expressed as couplings between band powers.  In
the case of instrument errors, we have seen that the visibility errors can
be expressed as linear combinations of the visibilities themselves.
In other words, the Fourier-space window functions associated with
the errors have exactly the same shape as the visibilities themselves.
In the case of beam errors, this is not strictly true: $\delta V_Q$,
for instance, has a different window function from $V_Q$.  However, for
all of the errors considered in this paper, differences in Fourier space
sensitivity introduced by the errors are relatively small: in all cases,
the errors sample regions of Fourier space centered near ${\bf k}=2\pi{\bf u}$
with widths $\Delta u \sim \sigma^{-1}$, just as the visibilities themselves
do.  In short, the errors do not couple greatly different angular
scales to each other.  This contrasts with single-dish
imaging experiments, in which scale-scale coupling induced by
systematic errors is an important
consideration \cite{HHZ}.  

The results above were calculated using a simple and relatively conservative
model for propagating errors from visibilities to power spectrum
estimates.  In a real data set, each resolution element in the
Fourier plane would be sampled by multiple visibilities rather than
just one pair.  If we can assume that the errors in all of
these visibilities are independent of each other and have ``nice''
probability distributions (particularly that the errors are
centered on zero), then the estimates should be reduced by
a factor of $\sqrt N$ where $N$ is the number of independent
visibility pairs $(V_Q,V_U)$ per resolution element.  However, since
systematic errors often do not have nice statistical properties,
a more conservative approach may be warranted.  Even if errors
do not need to be {\it removed} to the levels indicated here, 
it seems safe to say that their properties need to be {\it studied} 
down to this level of precision in order to have confidence
in the results.

Although there is expected to be no cosmological circular polarization
in the CMB, it is worthwhile to consider the effects of circular polarization
in the context of systematic errors.  On the one hand, various
errors can couple any intrinsic
circular polarization that does exist (e.g., from foregrounds) 
into the linear polarization channels, resulting in spurious
$E$ and $B$ signals.  On a more positive note, assuming that there is
no intrinsic
circular polarization, monitoring the circular polarization visibilities
$V_V$ may provide a way to assess systematic errors.  In particular, 
in a linear experiment coupling errors (including polarizer misalignments)
lead to a contribution to $V_V$ that is correlated with the temperature
anisotropy.  Considering the level of control of these errors
that is required in a $B$ mode experiment, such a diagnostic may prove
quite useful.

\begin{acknowledgments}
I thank Andrei Korotkov, Peter Timbie, Carolina Calderon, and Greg Tucker
for valuable insights, and
the
Brown University Physics Department for its hospitality during the 
completion of this work.  
This work is supported by NSF grant 0507395 and NASA grant NNG04GI15G.
\end{acknowledgments}

\appendix*
\section{}

In Sections \ref{sec:insterr} and \ref{sec:beamresults}, 
we made use of the following
fact:  Let ${\bf v}$ be a complex Gaussian random vector with
mean zero and
covariance matrix
\beq
{\bf M}\equiv\langle{\bf v}\cdot{\bf v}^\dag\rangle,
\eeq
and let ${\bf A}$ be an arbitrary hermitian matrix.  Let $q$
be the quadratic form
\beq
q\equiv{\bf v}^\dag\cdot{\bf A}\cdot{\bf v}.
\eeq
Then the mean-square value of $q$ is
\beq
\langle q^2\rangle={\rm Tr}[({\bf A}\cdot{\bf M})^2]
+[{\rm Tr}({\bf A}\cdot{\bf M})]^2.
\eeq
This Appendix, which no one will ever read, provides a proof of this fact.

First, note that we can always reduce the problem
to an equivalent one in which ${\bf M}$ is the identity
matrix.  To see this, let ${\bf Q}$ be a matrix such that ${\bf M}={\bf Q}\cdot
{\bf Q}^\dag$ (e.g., by Cholesky decomposition).  Let
${\bf v}'={\bf Q}^{-1}{\bf v}$ and ${\bf A}'={\bf Q}^{\dag}\cdot
{\bf A}\cdot{\bf Q}$.  Then $q={\bf v}^{\prime\dag}\cdot{\bf A}'\cdot
{\bf v}'$ and $\langle {\bf v}'\cdot{\bf v}^{\prime\dag}\rangle$
is the identity matrix.
We will assume that this transformation has been made and drop the primes.

Now diagonalize the hermitian matrix ${\bf A}$:
\beq
{\bf A}={\bf R}^\dag\cdot\bm{\Lambda}\cdot{\bf R},
\eeq
where $\bm{\Lambda}$ is diagonal with real entries $\lambda_i$,
and ${\bf R}$ is unitary.
Let ${\bf v}'={\bf R}\cdot{\bf v}$.  The covariance
matrix of ${\bf v}'$ is the identity matrix:
\beq
\langle{\bf v}'\cdot{\bf v}^{\prime\dag}\rangle
={\bf R}\cdot\langle{\bf v}\cdot{\bf v}^\dag\rangle\cdot{\bf R}^\dag
={\bf R}\cdot{\bf R}^\dag={\bf 1}.
\eeq
We have
\beq
q={\bf v}^{\prime\dag}\cdot\bm{\Lambda}\cdot{\bf v}'
=\sum_i\lambda_i |v'_i|^2,
\eeq
and therefore
\beq
\langle q^2\rangle=\sum_{i,j}\lambda_i\lambda_j\langle
|v'_i|^2|v'_j|^2\rangle.
\eeq
Each of the quantities $v'_i$ is an independent complex Gaussian random
variable with mean zero and variance one, so
\beq
\langle|v'_i|^2|v'_j|^2\rangle=
\begin{cases}
1 & \mbox{if }i\ne j\\
2 & \mbox{if }i=j
\end{cases}.
\eeq
(For real numbers, the $i=j$ case would be 3 rather than 2.)

Writing this as $1+\delta_{ij}$, we conclude that
\begin{subequations}
\begin{align}
\langle q^2\rangle&=\sum_{i,j}\lambda_i\lambda_j+
\sum_{i,j}\lambda_i\lambda_j\delta_{ij}\\
&=\Bigl(\sum_i\lambda_i\Bigr)^2+\sum_i\lambda_i^2\\
&=[{\rm Tr}(\bm{\Lambda})]^2+{\rm Tr}(\bm{\Lambda}^2).
\end{align}
\end{subequations}
Since traces
are unchanged under similarity transformations,
${\rm Tr}(\bm{\Lambda})={\rm Tr}({\bf A})$ and ${\rm Tr}(\bm{\Lambda}^2)
={\rm Tr}({\bf A}^2)$.
We have thus established
the desired result.

\bibliography{polsys}

\end{document}